    \documentclass[reprint,
    superscriptaddress,
    amsmath,amssymb,
    aps,
    pre,
    longbibliography,
    ]{revtex4-2}

\usepackage{graphicx}% Include figure files
\usepackage{dcolumn}% Align table columns on decimal point
\usepackage{bm}% bold math
\usepackage{hyperref}% add hypertext capabilities

\newcommand{\kk}{\mathbf{k}}
\newcommand{\UF}{\mathbf{U}}
\newcommand{\dif}{\mathrm{d}}

\newcommand{\rr}{\mathbf{r}}
\newcommand{\vv}{\mathbf{v}}
\newcommand{\uu}{\mathbf{u}}

\newcommand{\oo}{\boldsymbol{\omega}}

\newcommand{\dt}{{d_t}}
\newcommand{\dr}{{d_r}}

\newcommand{\Dert}{\mathcal{D}_t}
\newcommand{\Tt}{T_t}
\newcommand{\Trot}{T_r}

\newcommand{\een}{\alpha}
\newcommand{\eet}{\beta}

\def\bal#1\eal{\begin{align}#1\end{align}}
\newcommand\BEQ{\begin{equation}}
\newcommand{\beq}{\begin{equation}}
\newcommand\EEQ{\end{equation}}
\newcommand{\eeq}{\end{equation}}
\newcommand\beQa{\begin{eqnarray}}
\newcommand\eeQa{\end{eqnarray}}
\newcommand{\nn}{\nonumber\\}

\newcommand{\tr}{t}

\newcommand{\hcs}{H}

\newcommand{\acell}{a_{\text{cell}}}
\newcommand{\Ncell}{N_{\text{cell}}}
\newcommand{\Lcell}{L_{\text{cell}}}
\newcommand{\scell}{\sigma_{\text{cell}}}
\def\a{{\alpha }}
\def\b{{\beta }}
\def\bt{\widetilde\b}
\def\at{\widetilde\a}
\begin{document}

%\preprint{APS/123-QED}

\title{Hydrodynamics of granular gases of inelastic and rough hard disks or spheres. II. Stability analysis}% Force line breaks with \\

\author{Alberto Meg\'ias}
 \email{albertom@unex.es}
 \affiliation{Departamento de F\'isica, Universidad de Extremadura, E-06006 Badajoz, Spain}

\author{Andr\'es Santos}%
 \email{andres@unex.es}
 \affiliation{Departamento de F\'isica, Universidad de Extremadura, E-06006 Badajoz, Spain}
\affiliation{%
Instituto de Computaci\'on Cient\'ifica Avanzada (ICCAEx), Universidad de Extremadura, E-06006 Badajoz, Spain
}%

\date{\today}% It is always \today, today,

\begin{abstract}
Conditions for the stability under linear perturbations around the homogeneous cooling state  are studied for dilute granular gases of inelastic and rough hard disks or spheres with constant coefficients of normal ($\alpha$) and tangential ($\beta$) restitution. After a formally exact linear stability analysis of the Navier--Stokes--Fourier hydrodynamic equations in terms of the translational ($d_t$) and rotational ($d_r$) degrees of freedom, the transport coefficients derived in the companion paper [A.\ Meg\'ias and A.\ Santos, ``Hydrodynamics of granular gases of inelastic and rough hard disks or spheres. I. Transport coefficients,'' Phys. Rev. E \textbf{104}, 034901 (2021)] are employed. Known results for hard spheres [V.\ Garz\'o, A. Santos, and G. M. Kremer,  Phys.\ Rev.\ E \textbf{97}, 052901 (2018)] are recovered by setting $d_t=d_r=3$, while novel results for hard disks ($d_t=2$, $d_r=1$) are obtained. In the latter case,  a high-inelasticity  peculiar region in the $(\alpha,\beta)$ parameter space is found, inside which the critical wave number associated with the longitudinal modes diverges. Comparison with  event-driven molecular dynamics simulations for dilute systems of hard disks at $\alpha=0.2$ shows  that this theoretical region of absolute instability may be an artifact of the extrapolation to high inelasticity of the approximations made in the derivation of the transport coefficients, although it signals a shrinking of the conditions for stability. In the case of moderate inelasticity ($\alpha=0.7$), however, a good agreement between the theoretical predictions and the simulation results is found.
\end{abstract}

\maketitle

%\tableofcontents

%\section{\label{sec:level1}First-level heading:\protect\\ The line break was forced \lowercase{via} \textbackslash\textbackslash}

\section{Introduction}\label{sec:1}

Hard disks and spheres are very common models for describing fluids. In the molecular case, energy is conserved upon collisions, which are set to be elastic. As a consequence, the equilibrium state is obviously stable.
In contrast, a distinctive feature of a granular gas, as compared to a common fluid, is the possible instability of spatially uniform states and the associated appearance of structure formations (clusters and vortices). Characterization of the spontaneous formation of these instabilities has been widely studied for granular gases modeled as inelastic but smooth particles  \cite{GZ93,M93b,MY94,MY96,BDKS98,LH99,FH17,G19}. In these systems, it is always possible to find a range of parameters and perturbation wave numbers where a hydrodynamic description holds and instabilities are suppressed. In  recent years, this study has been expanded to the case of rough spheres \cite{MDHEH13,GSK18}, where a  dual role of roughness on instability has been observed.

Structure phenomena are important and appealing from a physical point of view. At a cosmological level, whereas the universe is considered to be generally isotropic and homogeneous, clustering is essential to forge galaxies and is present in planetary systems, dust agglomerations, planet rings, etc.  Moreover, vortex formation can remind the rotational motion of disk or spiral galaxies like our Milky Way.
From that point of view, granular gases, apart from their intrinsic interest, can serve as useful examples for the formation of clusters and vortices. However,  whereas attractive gravitational forces are the key of the clustering in the universe, in a granular gas the inelastic nature of the interacting particles is enough to produce it. Even more, the instabilities in self-gravitating granular gas systems has also been recently studied \cite{K20}. A similarity aspect between both classes of systems is that in cosmology one needs primordial perturbations in the early universe for the formation of agglomerations, while in a granular gas one can observe cluster formation spontaneously due to the growth of a given long enough perturbation. Furthermore, granular friction effects are known to have an influence on some astronomical problems \cite{BKBSHSS15,BRC17}.

In this paper, we consider a dilute granular gas modeled as a collection of hard spheres (HS) or hard disks (HD) which collide with constant coefficients of normal ($\alpha$) and tangential ($\beta$) restitution; while $0<\alpha\leq 1$ controls the degree of inelasticity, $-1\leq \beta\leq 1$ measures the degree of surface roughness. In general, each particle is animated with $\dt$ components of the translational velocity $\vv$ and $\dr$ components of the angular velocity $\oo$, where $(\dt,\dr)=(3,3)$ and $(2,1)$ for HS and HD, respectively. Our main aim is to perform a linear stability analysis of the homogeneous cooling state (HCS) of the granular gas by means of a Navier--Stokes--Fourier (NSF) hydrodynamic description in terms of the number of translational ($\dt$) and rotational ($\dr$) degrees of freedom, thus  encompassing the HS and HD systems within a unified treatment, as done in previous works \cite{MS19,MS19b,paperI}. To that end, we make explicit use of the approximate expressions for the NSF transport coefficients derived in the companion paper I \cite{paperI}. The HS results \cite{GSK18} are recovered by setting $(\dt,\dr)=(3,3)$, while novel results, to the best of our knowledge, are presented for HD by the choice $(\dt,\dr)=(2,1)$. In the latter case, we additionally present event-driven molecular dynamics (MD) simulations, where the possible emergence of instability is monitored via a coarse-grained Kullback--Leibler divergence (KLD) \cite{KL51,K78}, which measures the degree of spatial heterogeneities, as well as by the evolution of other relevant quantities (temperature ratio and velocity cumulants). As we will see, our results do not confirm previous studies \cite{PDR14}, where different cooling laws for rotational and translational temperatures were reported.

The paper is structured as follows. In Sec.\ \ref{sec:2}, the NSF hydrodynamic equations are presented for a granular gas in terms of the translational ($\dt$) and rotational ($\dr$) degrees of freedom. Afterwards, the linear stability analysis of the $\dt+2$ hydrodynamic equations around the HCS is completed in Sec.\ \ref{sec:3} in a formally exact way, that is, without assuming any particular form for the NSF transport coefficients. Next, in Sec.\ \ref{sec:4}, use is made of the approximate transport coefficients computed in Ref.\ \cite{paperI} and the results of the stability analysis are discussed. To clarify some unexpected outcomes in the HD case, our MD simulation results are exposed in Sec.\ \ref{sec:5}. Finally, concluding remarks of the work are presented in Sec.\ \ref{sec:6}.

\section{Navier--Stokes--Fourier hydrodynamic equations}\label{sec:2}

Let us consider a dilute granular gas made of identical HD ($\dt=2$, $\dr=1$) or HS ($\dt=\dr=3$) of diameter $\sigma$, mass $m$, and moment of inertia $I=\kappa m\sigma^2/4$, where $\kappa$ is the reduced moment of inertia. As said before, the collision dynamics will be assumed to be governed by two constant coefficients of restitution: normal ($\een$) and tangential ($\eet$).
In a kinetic-theory description of the gas, the mesoscopic relevant quantity is the one-body velocity distribution function $f(\rr,\vv,\oo;t)$, which obeys the Boltzmann equation.

At a macroscopic level, the adopted hydrodynamic fields are the number density $n(\rr,t)$, the flow velocity $\uu(\rr,t)$, and the temperature $T(\rr,t)$, which are defined as
\begin{subequations}
\beq
n(\rr,t)=\int \dif\vv\int \dif\oo\,f(\rr,\vv,\oo;t),
\eeq
\beq
\uu(\rr,t)=\frac{\int \dif\vv\int \dif\oo\,\vv f(\rr,\vv,\oo;t)}{n(\rr,t)},
\eeq
\beq
T(\rr,t)=\frac{\int \dif\vv\int \dif\oo\,\left\{m\left[\vv-\uu(\rr,t)\right]^2+I\omega^2\right\} f(\rr,\vv,\oo;t)}{(\dt+\dr)n(\rr,t)}.
\eeq

\end{subequations}

By assuming a Chapman--Enskog expansion around the HCS, the hydrodynamic equations to first order in the hydrodynamic gradients (NSF order) become
\begin{subequations}
\label{eq:17-120}
       \BEQ\label{eq:17}
        \Dert n =-n \nabla\cdot\uu ,
        \EEQ
  \bal
  \label{eq:120a}
    mn\Dert u_i =& -\tau_\tr \nabla_i(nT)+\nabla_j\bigg[\eta(\nabla_i u_j+\nabla_j u_i) \nn
    & -\left(\frac{2}{\dt}\eta-\eta_b\right)\delta_{ij}\nabla\cdot\uu \bigg],
\eal
    \bal\label{eq:120b}
    (\Dert+\zeta^{(0)})T =& \left(\xi-\frac{2\tau_\tr}{\dt+\dr} \right)T\nabla\cdot\uu \nn
    &+\frac{2}{(\dt+\dr)n}\nabla\cdot(\lambda\nabla T+\mu\nabla n)\nn
    &+\frac{2}{(\dt+\dr)n}\bigg[\eta\left(\nabla_i u_j+\nabla_j u_i \right)\nn
    &-\left(\frac{2}{\dt}\eta-\eta_b \right)\delta_{ij}\nabla\cdot\uu\bigg]\nabla_i u_j.
 \eal
    \end{subequations}
In these equations, $\Dert=\partial_t+\uu\cdot\nabla$ is the material time derivative,  $\tau_t$ is the HCS translational-to total temperature ratio, $\zeta^{(0)}$ is the Euler-order cooling rate, $\eta$ is the shear viscosity, $\eta_b$ is the bulk viscosity, $\lambda$ is the thermal conductivity, $\mu$ is a Dufour-like transport coefficient, and $\xi$ is a dimensionless transport coefficient associated with the velocity-divergence contribution to the cooling rate \cite{paperI}.

Dimensional analysis dictates that $\zeta^{(0)}=\zeta^*\nu$, $\eta=\eta^*\eta_0$, $\eta_b=\eta_b^*\eta_0$, $\lambda=\lambda^*\lambda_0$, and $\mu=\mu^* \lambda_0 T/n$, where
\BEQ\label{eq:40}
    \nu = K  n \sigma^{\dt-1}\sqrt{\frac{2\tau_tT}{m}}, \quad K\equiv \frac{\sqrt{2}\pi^{\frac{\dt-1}{2}}}{\Gamma\left(\dt/2\right)},
\EEQ
is a collision frequency, and
\beq
\eta_0=K_\ell \frac{n\tau_tT}{\nu}, \quad \lambda_0=\frac{2\dt K_\ell^2}{\dt-1} \frac{n\tau_tT}{m\nu},\quad K_\ell\equiv\frac{\dt+2}{4},
\eeq
are the shear viscosity and thermal conductivity, respectively, of a gas of elastic ($\alpha=1$) and smooth ($\beta=-1$) particles.
Apart from that, the explicit forms of the dimensionless coefficients $\tau_t$, $\zeta^*$, $\xi$, $\eta^*$, $\eta_b^*$, $\lambda^*$, and $\mu^*$ will not be needed for the moment.

\section{Linear Stability Analysis of the Homogeneous Cooling State}\label{sec:3}
The set of hydrodynamic equations given by Eqs.\ \eqref{eq:17-120} admits  the HCS as a special solution, in which $\nabla\to 0$ and thus the right-hand sides vanish. In that case, $n_\hcs=\text{const}$, $\uu_\hcs=\text{const}$, and $\dot{T}_\hcs=-\zeta^*\nu_\hcs T_\hcs$, where the quantities in the HCS are denoted with the subscript $\hcs$. Thus, $\nu\to\nu_\hcs\propto n_\hcs\sqrt{T_\hcs}$, $\eta_0\to \eta_{0\hcs}\propto \sqrt{T_\hcs}$, and $\lambda_0\to  \lambda_{0\hcs}\propto \sqrt{T_\hcs}$.
Moreover, we introduce the thermal (translational) velocity in the HCS as $v_\hcs=\sqrt{2\tau_tT_\hcs/m}$.

In this section we study the stability of the HCS by means of a linear perturbation analysis of the NSF equations, Eqs.\ \eqref{eq:17-120}. This study is essential to characterize the well-known structure formation that appears in granular gases.

The perturbations  of the hydrodynamic fields around the HCS are written as
\begin{subequations}
\label{delta y}
\beq
    n(\rr,t) = n_\hcs + \delta n(\rr,t),  \quad \uu(\rr,t) = \delta\uu(\rr,t),
\eeq
\beq
     T(\rr,t) = T_\hcs + \delta T(\rr,t),
\eeq
\end{subequations}
where, without loss of generality, we have chosen a reference frame with $\uu_\hcs=0$. By inserting Eqs.\ \eqref{delta y} into Eqs.\ \eqref{eq:17-120}, and neglecting terms nonlinear in the perturbations, we find
\begin{subequations}
\label{eq:122a-c}
 \beq
    \partial_t \frac{\delta n}{n_\hcs} = -v_\hcs \nabla\cdot\frac{\delta\uu}{v_\hcs}, \label{eq:122a}
 \eeq
 \bal
    \partial_t\frac{\delta u_i}{v_\hcs} =& \frac{\zeta^*\nu_\hcs}{2}\frac{\delta u_i}{v_\hcs}-\frac{v_\hcs}{2}\nabla_i\left(\frac{\delta n}{n_\hcs}+ \frac{\delta T}{T_\hcs} \right)
    +\frac{\eta_{0\hcs}}{m n_\hcs}\nn
    &\times\left[\left(\frac{\dt-2}{\dt}\eta^{*}+\eta_b^{*}\right)\nabla_i\nabla\cdot\frac{\delta\uu}{v_\hcs}+\eta^{*}\nabla^2\frac{\delta u_i}{v_\hcs}  \right], \label{eq:122b}
 \eal
 \bal
    \partial_t \frac{\delta T}{T_\hcs} =& -\zeta^*\nu_\hcs\left(\frac{\delta n}{n_\hcs}+\frac{\delta T}{2T_\hcs} \right)+\left(\xi-\frac{2\tau_\tr}{\dt+\dr} \right)v_\hcs\nn
    &\times \nabla\cdot \frac{\delta\uu}{v_\hcs}+\frac{2\lambda_{0\hcs}}{(\dt+\dr)n_\hcs}\nabla^2\left(\lambda^{*}\frac{\delta T}{T_\hcs}+\mu^{*}\frac{\delta n}{n_\hcs} \right).\label{eq:122c}
 \eal
\end{subequations}
Equations \eqref{eq:122a-c} form a closed set of $\dt+2$ linear partial differential equations.

It is now convenient to introduce the following scaled time and space variables,
\BEQ
    s(t) = \frac{1}{2}\int_0^t\dif t^\prime\, \nu_\hcs(t^\prime), \quad \boldsymbol{\ell} =\frac{\nu_\hcs}{\sqrt{2}K_\ell v_\hcs}\rr.
\EEQ
The variable $s$ measures the average number of collisions per particle, while $\boldsymbol{\ell}$ represents distance in units of a nominal mean free path (note that $\nu_\hcs/v_\hcs$ is independent of time).
Given a perturbation field $\delta y(\rr,t)$, we define its Fourier transform as
\BEQ
    \delta\widetilde{y}_\kk(s) = \int\dif\boldsymbol{\ell}\, e^{-\imath\kk\cdot\boldsymbol{\ell}}\delta y(\rr,t),
\EEQ
where $\imath$ is the imaginary unit and $\kk$ is the reduced wave vector. Thus, by defining the dimensionless quantities
\BEQ
    \rho_\kk(s) = \frac{\delta \widetilde{n}_\kk(s)}{n_\hcs}, \quad \UF_\kk(s) = \sqrt{2}\frac{\delta \widetilde{\uu}_\kk(s)}{v_\hcs}, \quad \Theta_\kk(s) = \frac{\delta\widetilde{T}_\kk(s)}{T_\hcs},
\EEQ
and taking the Fourier transform of Eqs.\ \eqref{eq:122a-c}, we obtain
\begin{subequations}
\beq
\label{rho_k}
 K_\ell \partial_s \rho_\kk = -\imath\kk\cdot\UF_\kk,
\eeq
\bal
\label{Uk}
 K_\ell \partial_s \UF_{\kk} =&\left(K_\ell\zeta^{*}-\frac{\eta^{*}k^2}{2}\right)\UF_{\kk}-\bigg[\imath\left(\Theta_\kk+\rho_\kk \right)\nn
 &+\frac{1}{2}\left(\frac{\dt-2}{\dt}\eta^{*}+\eta_b^{*} \right)\kk\cdot\UF_\kk \bigg]\kk,
\eal
\bal
\label{Theta_k}
 K_\ell\partial_s \Theta_\kk =& -K_\ell\zeta^{*}\left(2\rho_\kk+\Theta_\kk \right)+\imath\left(\xi-\frac{2\tau_\tr}{\dt+\dr}\right)\kk\cdot\UF_\kk\nn
 &-\frac{2\dt K_\ell}{(\dt-1)(\dt+\dr)} k^2(\lambda^{*}\Theta_\kk+\mu^{*}\rho_\kk).
\eal
\end{subequations}

Taking the inner product with $\kk$ in both sides of Eq.\ \eqref{Uk}, one gets
\bal
\label{Ukpar}
 K_\ell \partial_s U_{\kk,\|} =&\left(K_\ell\zeta^{*}-\frac{\eta^{*}k^2}{2}\right)U_{\kk,\|}-\bigg[\imath\left(\Theta_\kk+\rho_\kk \right)\nn
 &+\frac{1}{2}\left(\frac{\dt-2}{\dt}\eta^{*}+\eta_b^{*} \right)k U_{\kk,\|} \bigg]k,
\eal
where $U_{\kk,\|}=k^{-1}\kk\cdot\UF_\kk$ is the longitudinal component of the vector $\UF_\kk$. Next, combination of Eqs.\ \eqref{Uk} and \eqref{Ukpar} yields the following equation for the $\dt-1$ transverse components  $\UF_{\kk,\perp}=\UF_\kk-U_{\kk,\|}\kk/k$,
\BEQ
\label{Ukperp}
    \left(\partial_s- \zeta^{*}+\frac{\eta^{*}k^2}{2K_\ell}\right)\UF_{\kk,\perp}=0.
\EEQ
Thus, the transverse vector $\UF_{\kk,\perp}$ decouples from the other three hydrodynamic fields. The solution to Eq.\ \eqref{Ukperp} is simply
\BEQ
\label{w_perp}
    \UF_{\kk,\perp}(s) = \UF_{\kk,\perp}(0)e^{\varpi_\perp(k)s},\quad \varpi_\perp(k)=\zeta^{*}-\frac{\eta^{*}k^2}{2K_\ell}.
\EEQ
This characterizes the behavior of the $\dt-1$ \emph{shear} modes. They decay in time if $\varpi_\perp(k)<0$, i.e., if the (reduced) wave number $k$ is larger than a critical value
\BEQ\label{eq:132}
    k_\perp = \sqrt{\frac{2K_\ell \zeta^{*}}{\eta^{*}}}.
\EEQ
However, if $k<k_\perp$, then the shear modes grow in time and the HCS is unstable under those transverse perturbations.

We consider now the three longitudinal modes $\rho_\kk$, $\Theta_\kk$, and $U_{\kk,\|}$. Equations \eqref{rho_k}, \eqref{Theta_k}, and \eqref{Ukpar} can be rewritten in matrix form as
\beq
\partial_s
\begin{pmatrix}
  \rho_\kk\\
  \Theta_\kk\\
  U_{\kk,\|}
\end{pmatrix}
=\mathsf{M}\cdot
\begin{pmatrix}
  \rho_\kk\\
  \Theta_\kk\\
  U_{\kk,\|}
\end{pmatrix},
\eeq
where
\BEQ
    \mathsf{M}(k) =- \begin{pmatrix}
        0 & 0 & {\imath k}/{K_\ell}\\
  2\zeta^{*}+C_\mu k^2 & \zeta^{*}+C_\lambda k^2 & -C_\xi{\imath k}/{K_\ell}\\
     {\imath k}/{K_\ell} & {\imath k}/{K_\ell} & -\zeta^{*}+C_\eta{k^2}
    \end{pmatrix}.
\EEQ
Here,
\begin{subequations}
\beq
C_\lambda\equiv \frac{2\dt\lambda^{*}}{(\dt-1)(\dt+\dr)},\quad C_\mu\equiv \frac{2\dt\mu^{*}}{(\dt-1)(\dt+\dr)},
\eeq
\beq
C_\xi\equiv \xi-\frac{2\tau_\tr}{\dt+\dr},\quad C_\eta\equiv\frac{1}{K_\ell}\left(\frac{\dt-1}{\dt}\eta^{*}+\frac{\eta_b^{*}}{2}\right).
\eeq
\end{subequations}

Let us denote as $\varpi_{\|,1}(k)$, $\varpi_{\|,2}(k)$, and $\varpi_{\|,3}(k)$ the three eigenvalues of the matrix $\mathsf{M}$. They are given by the  roots of the  characteristic polynomial
\BEQ
\label{charact}
    \varpi_{\|}^3+F_2(k)\varpi_{\|}^2+F_1(k)\varpi_{\|}+F_0(k),
\EEQ
with
\begin{subequations}
\beq
    F_0(k)=\left[-\zeta^{*}+\left(C_\lambda-C_\mu\right)k^2\right]\frac{k^2}{K_\ell^2},
\eeq
\beq
    F_1(k)=-{\zeta^{*}}^2+\left[\frac{1-C_\xi}{K_\ell^2}+\left(C_\eta-C_\lambda\right)\zeta^*\right]k^2+C_\eta C_\lambda k^4,
\eeq
\beq
    F_2(k)= \left(C_\eta+C_\lambda\right)k^2.
\eeq
\end{subequations}

In the long-wavelength limit ($k\ll 1$), the roots of Eq.\ \eqref{charact} reduce to
\begin{subequations}
\beq
\varpi_{\|,1}(k)=-\zeta^*+\left(\frac{1-C_\xi/2}{K_\ell^2\zeta^*}-C_\lambda\right)k^2+\cdots,
\eeq
\beq
\varpi_{\|,2}(k)=-\frac{k^2}{K_\ell^2\zeta^*}+\cdots,
\eeq
\beq
\label{varpi3}
\varpi_{\|,3}(k)=\zeta^*-\left(C_\eta-\frac{C_\xi}{2K_\ell^2\zeta^*}\right)k^2+\cdots.
\eeq
\end{subequations}
The two eigenvalues $\varpi_{\|,1}$ and $\varpi_{\|,2}$ define a pair of \emph{sound} modes, while $\varpi_{\|,3}$ corresponds to the  \emph{heat} mode.
The heat mode is unstable for wave numbers ($k<k_{\|}$) such that $\varpi_{\|,3}(k)$ becomes positive. To determine the associated critical value $k_{\|}$, we set $\varpi_{\|}=0$ in Eq.\ \eqref{charact}, i.e., $F_0(k_{\|})=0$. Therefore,
\BEQ\label{eq:143}
    k_{\parallel} =\sqrt{\frac{(\dt-1)(\dt+\dr)}{2\dt}}\sqrt{\frac{\zeta^*}{\lambda^*-\mu^*}}.
\EEQ

{\renewcommand{\arraystretch}{2}
\begin{table}
   \caption{Summary of the explicit expressions of the transport coefficients for a granular gas of inelastic and rough HD in a Sonine-like approximation \cite{paperI}.}\label{table0}
\begin{ruledtabular}
\begin{tabular}{l}
$\displaystyle{\widetilde\a=\frac{1+\a}{2}}, \quad \displaystyle{\bt=\frac{1+\b}{2}\frac{\kappa}{1+\kappa}}$\\
$\displaystyle{\frac{\Tt^{(0)}}{T}=\tau_t=\frac{3}{2+\theta}}, \quad \displaystyle{\frac{\Trot^{(0)}}{T}=\tau_r=\frac{3\theta}{2+\theta}}$\\
$\displaystyle{\theta=\sqrt{\left(h-\frac{1}{2}\right)^2+2}+h-\frac{1}{2}}$\\
$\displaystyle{h\equiv \frac{(1+\kappa)^2}{\kappa(1+\b)^2}\left[{1-\a^2}-\frac{1-\frac{1}{2}\kappa}{1+\kappa}(1-\b^2)\right]}$\\
$\displaystyle{\nu=2 n\sigma\sqrt{\pi\tau_tT/m}}$\\
$\displaystyle{\frac{\zeta^{(0)}}{\nu}=\zeta^*=\frac{1}{2+\theta}\left[1-\a^2+\frac{1}{2}\frac{1-\b^2}{1+\kappa}
(\kappa+\theta)\right]}
$\\[2mm]
\hline
 $\displaystyle{\eta=\frac{n\tau_t T}{\nu}\frac{1}{\nu_\eta^*-\frac{1}{2}\zeta^*}},\quad \displaystyle{\eta_b=\frac{ n\tau_t\tau_r T}{2\nu}\gamma_E}$\\
$\displaystyle{\lambda={\tau_t\lambda_t+\tau_r\lambda_r}}, \quad \displaystyle{\lambda_t=\frac{2n\tau_t T}{m\nu}\gamma_{A_t}}, \quad\displaystyle{\lambda_r=\frac{n\tau_t T}{2m\nu}\gamma_{A_r}}$\\
 $\displaystyle{\mu={\mu_t+\mu_r}}, \quad\displaystyle{\mu_t=\frac{2\tau_t^2 T^2}{m\nu}\gamma_{B_t}}, \quad \displaystyle{\mu_r=\frac{\tau_t \tau_r T^2}{2m\nu}\gamma_{B_r}}$\\
$\displaystyle{\xi=\frac{1}{3}\left(2\tau_t\xi_t+\tau_r\xi_r\right)=\gamma_E\Xi}, \quad\displaystyle{\xi_t=\gamma_E\Xi_t}, \quad\displaystyle{\xi_r=\gamma_E\Xi_r}$\\[1mm]
\hline
$\displaystyle{\nu_\eta^*=\frac{5\at}{2}+\frac{5\bt}{4}-\frac{3\at^2}{2}-\frac{\bt^2}{4}-2\at\bt+\frac{\bt^2\theta}{4}}$\\
$\displaystyle{\gamma_E=\left({\Xi_t-\Xi_r-\frac{3}{4}\zeta^*}\right)^{-1}}$\\
$\displaystyle{\Xi_t=\frac{3\tau_r}{8}\left[1-\a^2+\frac{1}{2}\frac{\kappa}{1+\kappa}(1-\b^2)-\left(\frac{1+\b}{1+\kappa}\right)^2\kappa\frac{\theta-7}{6}
\right]}$\\
$\displaystyle{\Xi_r=\frac{\tau_t}{4}\frac{1+\b}{1+\kappa}\left[(1-\b)\frac{\theta-4}{2}
+\frac{1+\b}{1+\kappa}\kappa\frac{\theta-7}{2}\right]}$\\
$\displaystyle{\Xi=\frac{\tau_t\tau_r}{4}\left(1-\a^2+\frac{1-\b^2}{1+\kappa}\frac{3\kappa+\theta-4}{6}\right)}$\\
$\displaystyle{\gamma_{A_t}=\frac{Z_r-Z_t-2\zeta^*}{\left(Y_t-2\zeta^*\right)\left(Z_r-2\zeta^*\right)-Y_r Z_t}}$\\
$\displaystyle{\gamma_{A_r}=\frac{Y_t-Y_r-2\zeta^*}{\left(Y_t-2\zeta^*\right)\left(Z_r-2\zeta^*\right)-Y_r Z_t}}$\\
$\displaystyle{\gamma_{B_t}=\zeta^*\frac{\gamma_{A_t}\left(Z_r-\frac{3}{2}\zeta^*\right)-\gamma_{A_r}Z_t}
{\left(Y_t-\frac{3}{2}\zeta^*\right)\left(Z_r-\frac{3}{2}\zeta^*\right)-Y_r Z_t}}$\\
$\displaystyle{\gamma_{B_r}=\zeta^*\frac{\gamma_{A_r}\left(Y_t-\frac{3}{2}\zeta^*\right)-\gamma_{A_t}Y_r}
{\left(Y_t-\frac{3}{2}\zeta^*\right)\left(Z_r-\frac{3}{2}\zeta^*\right)-Y_r Z_t}}$\\
$\displaystyle{Y_t=\frac{17\at}{4}+\frac{17\bt}{8}-\frac{15\at^2}{4}-\frac{13\bt^2}{8}-\at\bt-\frac{3\bt^2\theta}{8\kappa}}$\\
$\displaystyle{Y_r=\frac{\bt}{\kappa}\left(1-\frac{3\bt}{\theta}-\frac{\bt}{\kappa}\right)},\quad \displaystyle{Z_t=-\frac{\bt^2\theta}{2\kappa}
}$\\
$\displaystyle{Z_r=\at+\frac{\bt}{2}+\frac{\bt}{\kappa}\left(\frac{5}{2}-2{\at}-2\bt-\frac{\bt}{\kappa}\right)}$\\
     \end{tabular}
 \end{ruledtabular}
 \end{table}
}

\begin{figure}
    \includegraphics[width=\columnwidth]{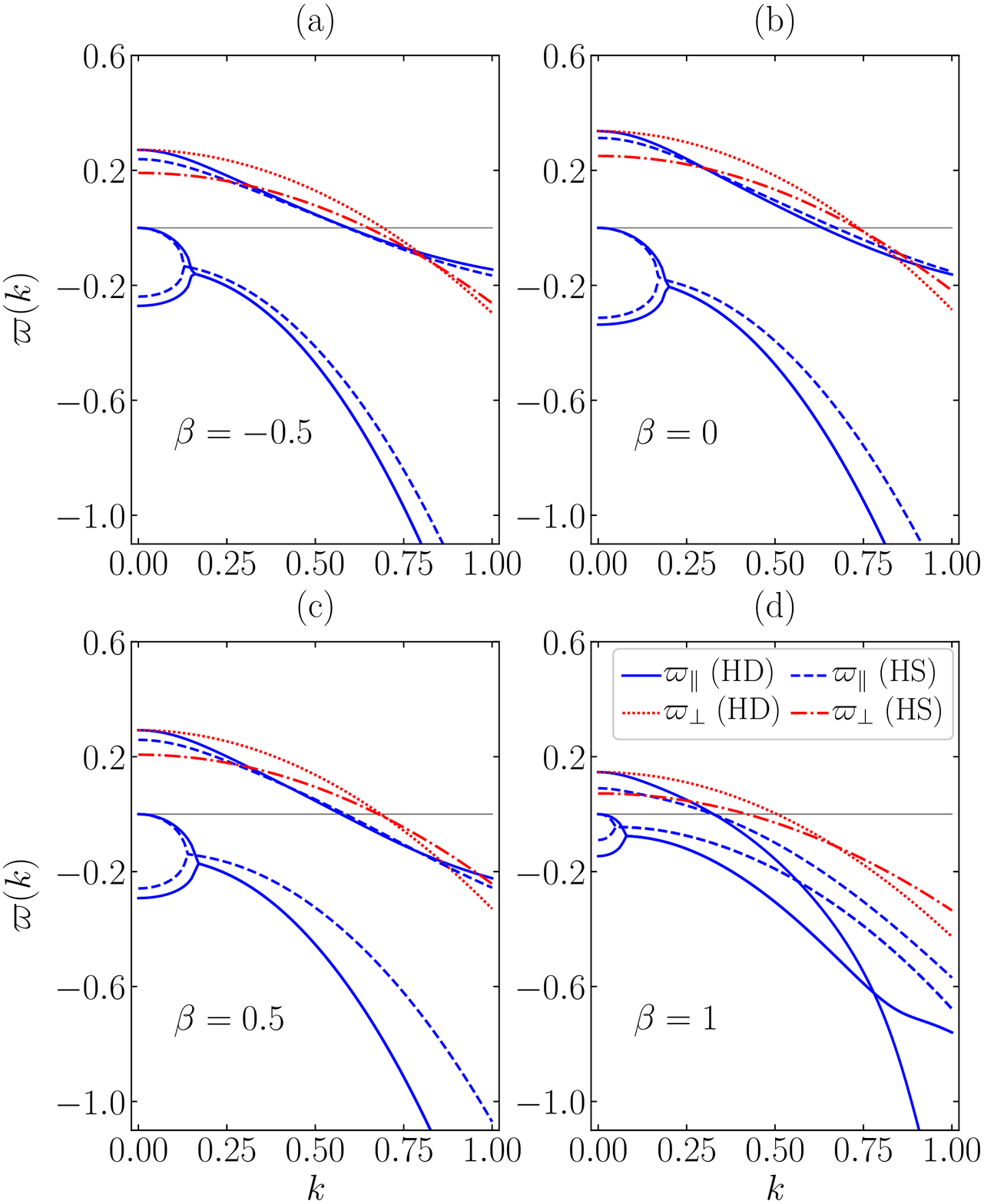}
    \caption{Dispersion relations $\varpi(k)$ for the hydrodynamic modes vs the reduced wave number $k$. The curves correspond to the degenerate shear mode $\varpi_\perp$ (red lines), the heat mode $\varpi_{\|,3}$, and the sound modes  $\varpi_{\|,1}$ and $\varpi_{\|,2}$ (blue lines). Note that when $\varpi_{\|,1}$ and $\varpi_{\|,2}$ become a complex conjugate pair, only the (common) real part is plotted.
The solid and dotted lines represent the HD system, while the dashed and dash-dotted lines refer to HS systems. The coefficient of normal restitution is $\alpha=0.7$, the reduced moment of inertia is $\kappa=\frac{1}{2}$ (HD) or $\kappa=\frac{2}{5}$ (HS), and the coefficients of tangential restitution are (a) $\beta=-0.5$, (b) $\beta=0$, (c) $\beta=0.5$, and (d) $\beta=1$.}
    \label{fig:w_07}
\end{figure}

\begin{figure}
    \includegraphics[width=\columnwidth]{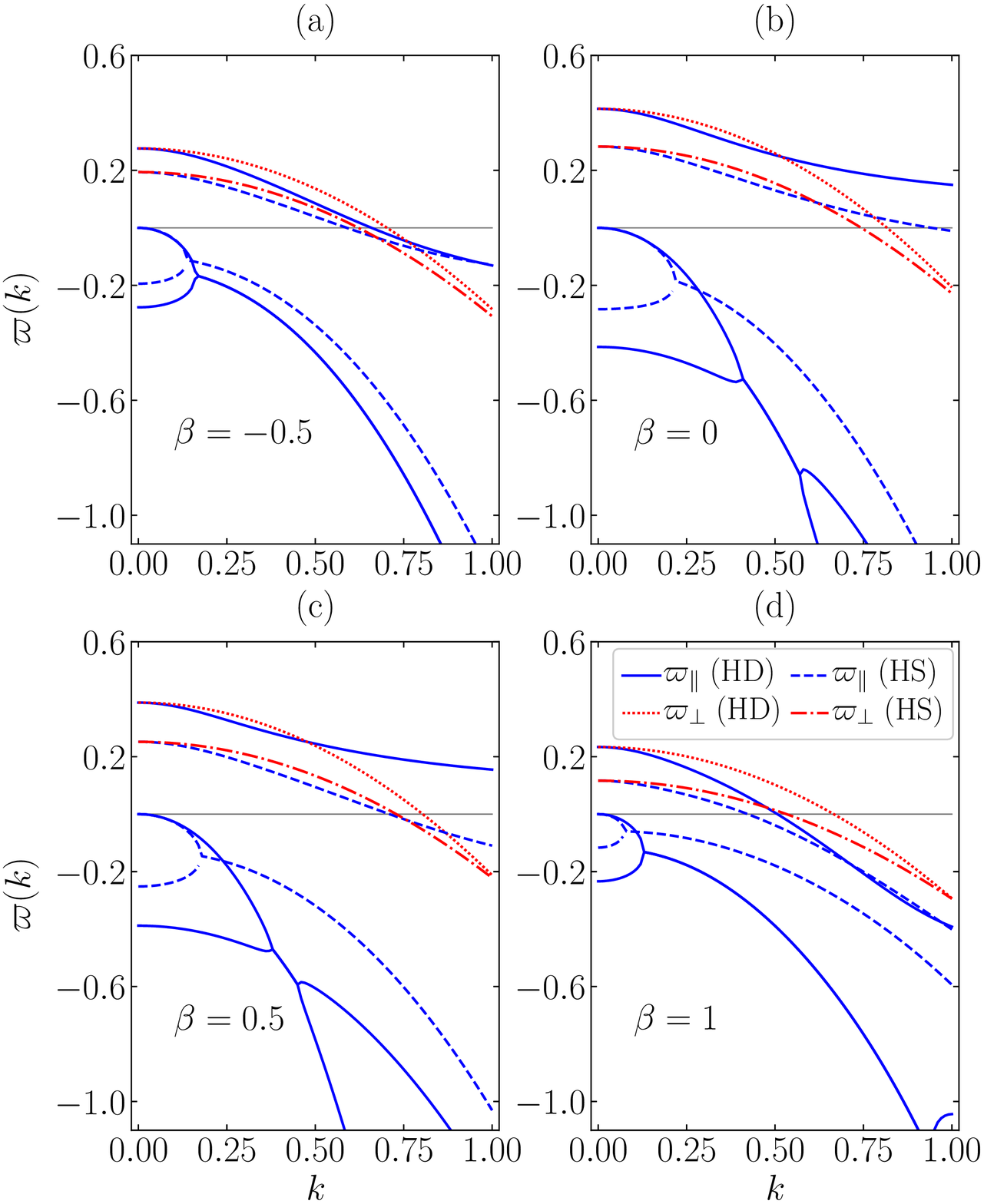}
    \caption{Same as described in the caption of Fig.\ \ref{fig:w_07}, except that $\alpha=0.2$.}
    \label{fig:w_02}
\end{figure}

\begin{figure}
    \includegraphics[width=\columnwidth]{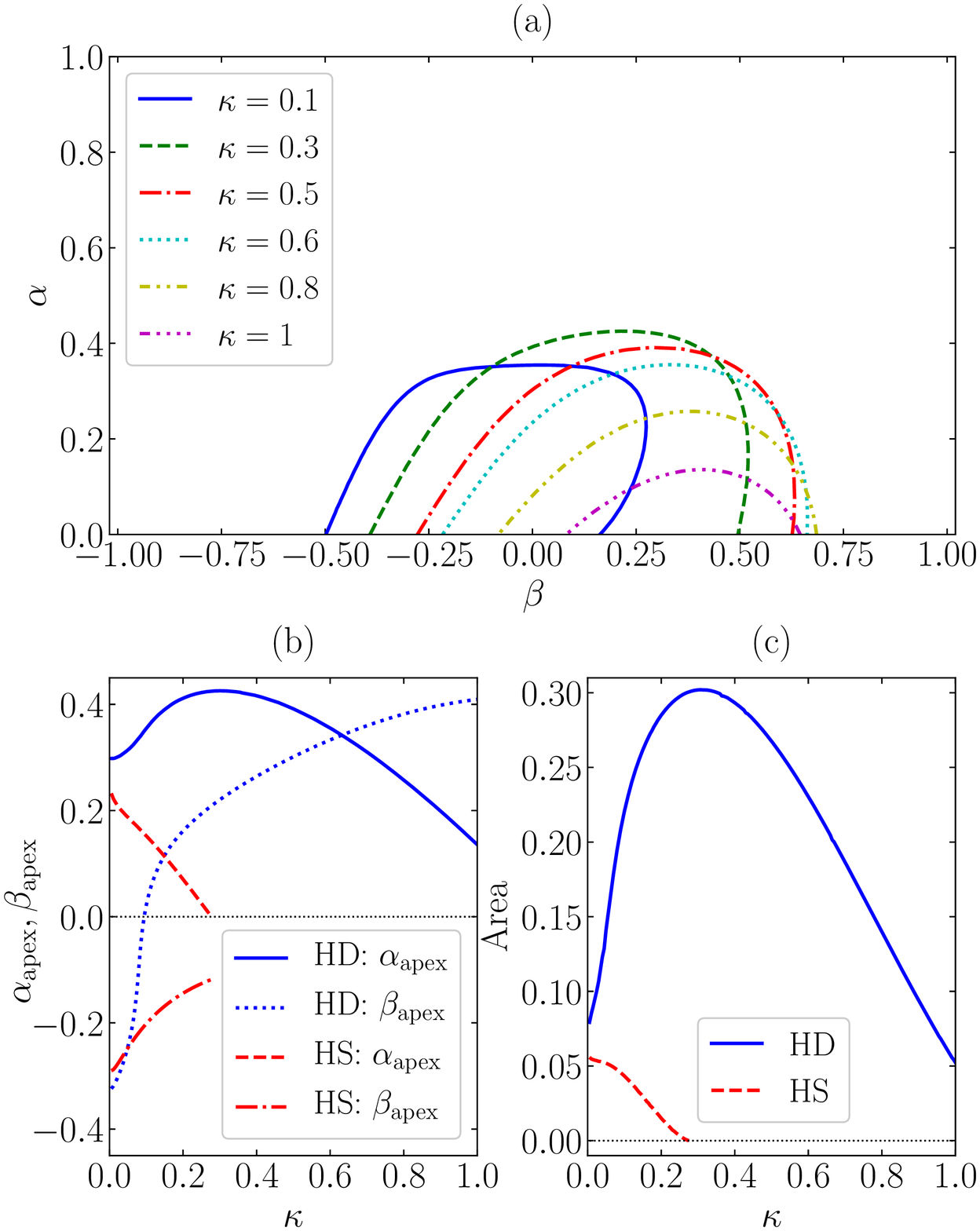}\\
    \caption{(a) Plane $\alpha$ vs  $\beta$ showing the locus $\lambda^*=\mu^*$ for HD with a reduced moment of inertia $\kappa=0.1$, $0.3$, $0.5$, $0.8$, and $1$. In each case, $k_\|\to\infty$ in the region below the locus, which has an apex located at $(\alpha,\beta)=(\alpha_{\text{apex}}, \beta_{\text{apex}})$.
    (b) Dependence  of $\alpha_{\text{apex}}$ and $\beta_{\text{apex}}$ on $\kappa$ for HD and HS. (c) Variation with $\kappa$ of the area of the region where $k_\|\to\infty$ for HD and HS.}
    \label{fig:Locus}
\end{figure}

\begin{figure}
    \includegraphics[width=\columnwidth]{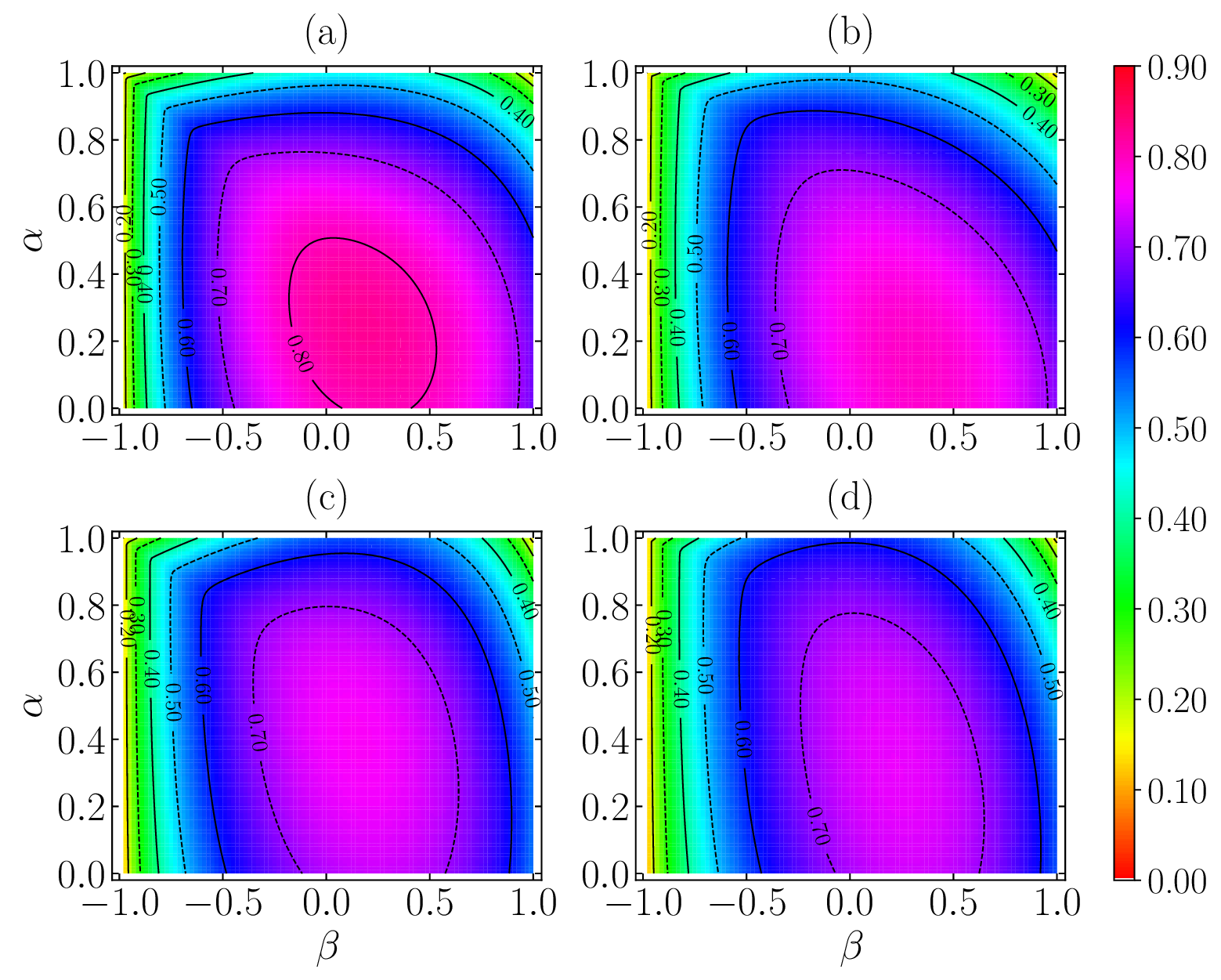}
    \caption{Density plots of the reduced critical wave number $k_\perp$ in the plane $\alpha$ vs $\beta$   for (a) HD with a uniform mass distribution ($\kappa=\frac{1}{2}$), (b) HD with a  mass distribution concentrated on the outer surface ($\kappa=1$), (c) HS with a uniform mass distribution ($\kappa=\frac{2}{5}$), and (d) HS with a  mass distribution concentrated on the outer surface ($\kappa=\frac{2}{3}$).}
    \label{fig:kperp}
\end{figure}

\begin{figure}
    \includegraphics[width=\columnwidth]{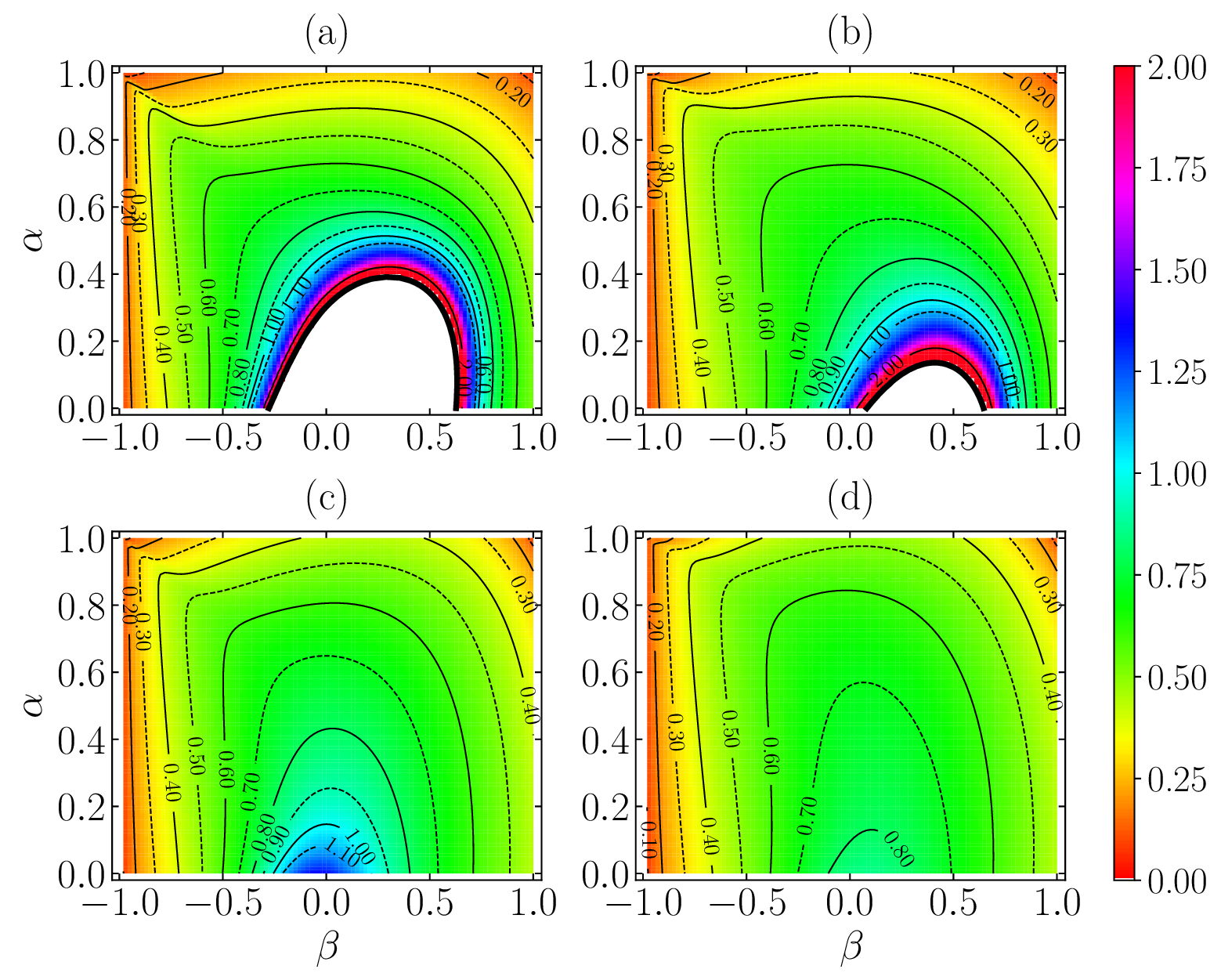}
    \caption{Same as described in the caption of Fig.\ \ref{fig:kperp}, but for the reduced critical wave number $k_\|$.}
    \label{fig:kpara}
\end{figure}

\begin{figure}
    \includegraphics[width=\columnwidth]{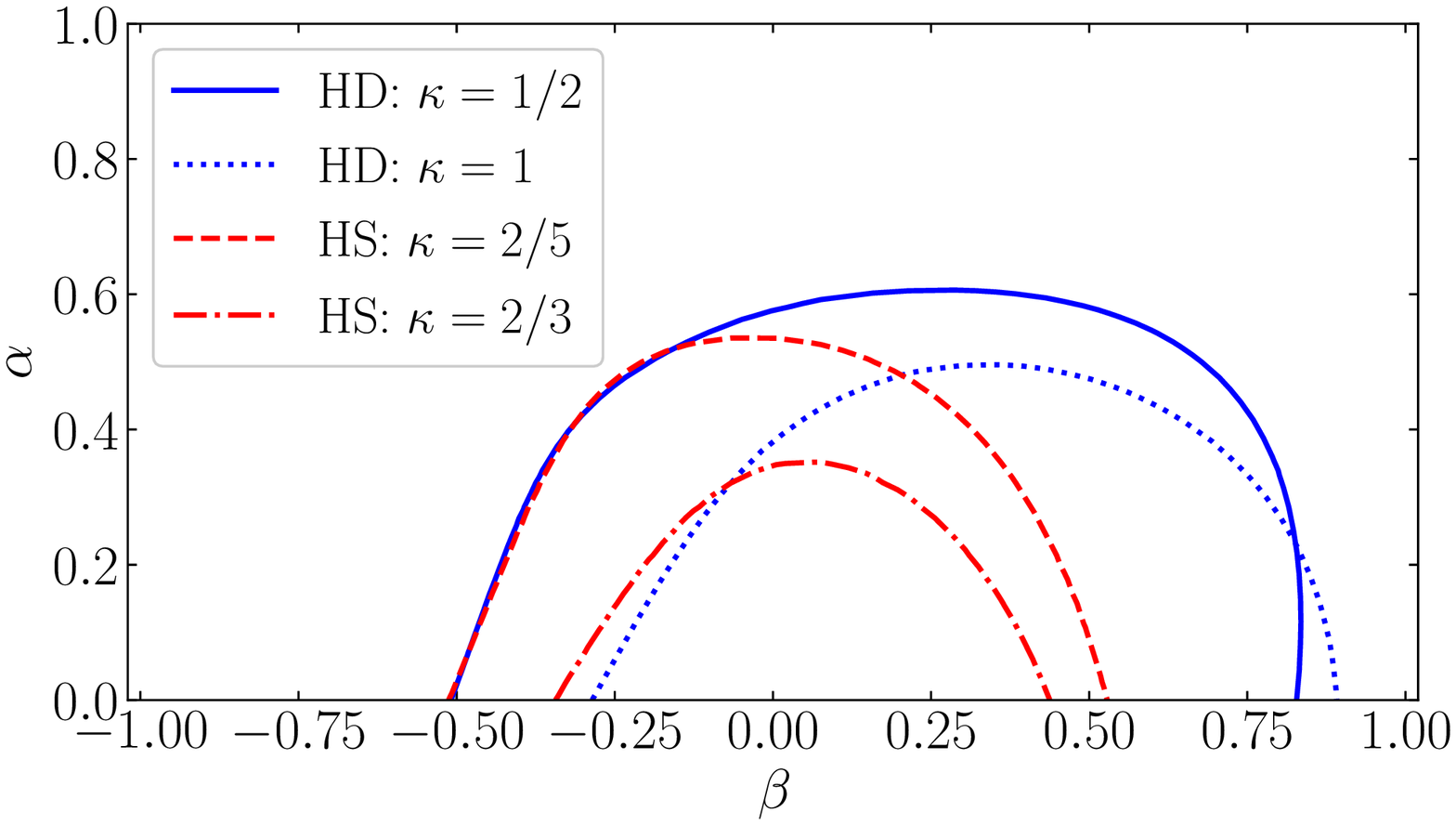}
    \caption{Plane $\alpha$ vs $\beta$  showing the locus $k_\perp=k_\|$ for HD ($\kappa=\frac{1}{2}$ and $1$) and HS ($\kappa=\frac{2}{5}$ and $\frac{2}{3}$). In each case, the longitudinal heat mode is the most unstable one ($k_\|>k_\perp$) in the region below the locus.}
    \label{fig:Locuskk}
\end{figure}

\section{Analysis}
\label{sec:4}

All the results in Secs.\ \ref{sec:2} and \ref{sec:3} are general in the sense that the explicit expressions for the dimensionless coefficients $\tau_t$,  $\zeta^*$, $\xi$, $\eta^*$, $\eta_b^*$, $\lambda^*$, and $\mu^*$ have not been used. Those coefficients are functions of the coefficients of restitution ($\alpha$, $\beta$) and the reduced moment of inertia ($\kappa$), and they also depend on the number of degrees of freedom ($\dt$ and $\dr$). As shown in paper I \cite{paperI}, the exact determination of $\tau_t$ and  $\zeta^*$ would require to solve the nonlinear Boltzmann equation for the zeroth-order HCS velocity distribution function $f^{(0)}$. The situation is even more involved in the case of the transport coefficients $\xi$, $\eta^*$, $\eta_b^*$, $\lambda^*$, and $\mu^*$, whose determination, assuming $f^{(0)}$ were already known, would require to solve four linear integral equations for the first-order distribution function $f^{(1)}$.

To overcome the above difficulties, in paper I we adopted a Sonine-like approximation for $f^{(1)}$ supplemented by a quasi-Maxwellian approximation for $f^{(0)}$ (see Ref.\ \cite{paperI} for details) that allowed us to obtain (approximate) explicit expressions for $\tau_t$,  $\zeta^*$, $\xi$, $\eta^*$, $\eta_b^*$, $\lambda^*$, and $\mu^*$ as functions of $\alpha$, $\beta$,  $\kappa$,  $\dt$, and $\dr$. The results are summarized in Table I of paper I and agree with those previously derived \cite{KSG14} for HS ($\dt=\dr=3$). For completeness, we present in Table \ref{table0} the results for HD ($\dt=2$, $\dr=1$), which, to the best of our knowledge, have not been shown before.

In the case of purely smooth particles ($\beta=-1$ or, in our approach, $\dr\to 0$), it is known that the HCS becomes unstable under perturbations with a sufficiently small wave number  ($k<\max\{k_\perp,k_\|\}$) \cite{BDKS98,G19}. Interestingly, the quasismooth limit $\beta\to -1$ is singular and yields $\zeta^*\to 0$ and $\mu^*\to 0$. Consequently, according to Eqs.\ \eqref{eq:132} and \eqref{eq:143}, $k_\perp,k_\|\to 0$. In the general case, however, the HCS of a granular gas  of rough HD or HS can be unstable.

Figure \ref{fig:w_07} shows the dispersion relations $\varpi(k)$, as obtained from Eqs.\ \eqref{w_perp} and \eqref{charact}, at $\alpha=0.7$ and for several representative values of $\beta$. In each case, uniform HD ($\kappa=\frac{1}{2}$) and uniform HS ($\kappa=\frac{2}{5}$) are considered. The curves for HD and HS are qualitatively similar. In both systems, the real part of the sound modes ($\varpi_{\|,1}$ and $\varpi_{\|,2}$) remain negative for all $k$, thus indicating that those perturbative modes decay in time. However, the shear ($\varpi_\perp$) and heat ($\varpi_{\|,3}$) modes grow in time if the wavenumber is smaller than $k_\perp$ and $k_\|$, respectively. Note that both frequencies ($\varpi_\perp$ and $\varpi_{\|,3}$) tend to $\zeta^*$ in the small wave number limit $k\to 0$ [see Eqs.\ \eqref{w_perp} and \eqref{varpi3}].

The scenario becomes much more complex for highly inelastic particles, as illustrated in Fig.\ \ref{fig:w_02} at $\alpha=0.2$. In the cases $\beta=-0.5$ [Fig.\ \ref{fig:w_02}(a)] and $\beta=1$ [Fig.\ \ref{fig:w_02}(d)], the HD and HS curves are still qualitatively similar. However, if $\beta=0$ [Fig.\ \ref{fig:w_02}(b)] or $\beta=0.5$ [Fig.\ \ref{fig:w_02}(c)], then $\varpi_{\|,3}>0$ for all $k$  (i.e., $k_\|\to\infty$) in the HD case. {}From Eq.\ \eqref{eq:143} we see that the locus in the plane $\alpha$ versus $\beta$  separating the region where $k_\|=\text{finite}$ from the region where $k_\|\to\infty$ is defined by the condition $\lambda^*=\mu^*$.

The locus $\lambda^*=\mu^*$ for HD is shown in Fig.\ \ref{fig:Locus}(a) for several values of $\kappa$. For each $\kappa$, $k_\|\to\infty$ in the  region enclosed by the locus. The latter curve presents an apex at a point $(\alpha,\beta)=(\alpha_{\text{apex}}, \beta_{\text{apex}})$, so that $k_\|=\text{finite}$ if $\alpha>\alpha_{\text{apex}}$, regardless of the value of $\beta$. A similar behavior occurs in the HS case \cite{GSK18}, except that the regions where $k_\|\to\infty$ are much smaller and disappear if $\kappa>0.277$.
The dependence of $\alpha_{\text{apex}}$ and $\beta_{\text{apex}}$ on $\kappa$ for both HD and HS is shown in Fig.\ \ref{fig:Locus}(b). While $\alpha_{\text{apex}}$ for HS decays monotonically as $\kappa$ increases (and eventually vanishes at $\kappa=0.277$),  it exhibits a nonmonotonic behavior for HD, with a maximum value $\alpha_{\text{apex}}=0.426$ at $\kappa=0.302$. However, $\beta_{\text{apex}}$ grows monotonically with $\kappa$ both for HD and HS. The contrast between the HD and HS behaviors is clearly highlighted in  Fig.\ \ref{fig:Locus}(c), which shows the $\kappa$-dependence of the area of the region where $k_\|\to\infty$.

Let us now visualize the dependence of the two critical wave numbers $k_\perp$ and $k_\|$ on $\alpha$, $\beta$, and $\kappa$ for both HD and HS systems. The results are shown as density plots in the plane $\alpha$  versus $\beta$ in Figs.\ \ref{fig:kperp} and \ref{fig:kpara}, where two representative mass distributions of the particles are considered: a uniform distribution ($\kappa=\frac{1}{2}$ and $\frac{2}{5}$ for HD and HS, respectively) and a distribution concentrated on the surface ($\kappa=1$ and $\frac{2}{3}$ for HD and HS, respectively). In the case of the transverse shear-mode critical wave number $k_\perp$, the dependence on $\alpha$, $\beta$, and $\kappa$ is qualitatively similar for HD and HS granular gases. However, this similarity disappears in what respects the longitudinal heat-mode critical wave number $k_\|$ as one approaches the HD locus $\lambda^*=\mu^*$, in agreement with the previous discussion of Fig.\ \ref{fig:Locus}.

Depending on the values of $\alpha$ and $\beta$, the most unstable mode could be either the transverse shear mode (if $k_\perp>k_\|$) or the longitudinal heat one (if $k_\|>k_\perp$). Figure \ref{fig:Locuskk} depicts the locus $k_\perp=k_\|$ for HD and HS gases and the same values of $\kappa$ as in Figs.\ \ref{fig:kperp} and \ref{fig:kpara}. In each case, $k_\perp>k_\|$ or $k_\|>k_\perp$ above or below the locus, respectively. We observe that the region where the heat mode dominates ($k_\|>k_\perp$) is generally wider for HD than for HS; moreover, its area  decreases as $\kappa$ increases for HS, while for HD it has a nonmonotonic $\kappa$-dependence  with a maximum at about $\kappa=0.348$ (not shown).

The critical wave numbers $k_\perp$ and $k_\|$ imply that the HCS becomes unstable if the (reduced) length of the system is larger than the critical value $\ell_c=2\pi/k_c$, where $k_c=\max\{k_\perp,k_\parallel \}$. In real units, the critical length is $L_c=(\sqrt{2}K_\ell v_\hcs/\nu_\hcs)\ell_c$, i.e.,
\beq
\frac{L_c}{\sigma}=\frac{(\dt+2)\pi^{3/2}}{\dt 2^{\dt}\phi}k_c^{-1},
\eeq
where
\BEQ
    \phi = \frac{\pi^{\dt/2}}{2^{\dt-1}\dt\Gamma(\dt/2)}n\sigma^{\dt}
\EEQ
is the solid fraction of the system.
At a given value of the reduced moment of inertia $\kappa$, $L_c$ is associated with either  vortex  or clustering instability in the region above or below, respectively, the corresponding locus in Fig.\ \ref{fig:Locuskk}. Moreover, $L_c\to 0$ in the region below the locus on Fig.\ \ref{fig:Locus}(a) for HD gases. In other words, in that region (henceforth referred to as the region of absolute instability), the HCS would always be unstable for any system size. This is a very strong statement that needs some discussion.

Note that the condition $\lambda^*\leq \mu^*$ takes place for very inelastic disks and never holds if $\alpha> 0.426$. Since the explicit expressions for the transport coefficients derived in paper I \cite{paperI} made use of a Sonine-like approximation for the first-order distribution $f^{(1)}$ and a quasi-Maxwellian approximation for the zeroth-order distribution $f^{(0)}$, it cannot be discarded that the combination of those two approximations is responsible for the existence of the region of absolute instability. If that were the case, then a more sophisticated approximation, for instance, by consistently including the cumulants of $f^{(0)}$ in the description, would erase such a  region and $\lambda^*$ would be larger than $\mu^*$ for any $\alpha$, $\beta$, and $\kappa$.

To put that possibility in context, let us recall the case of purely smooth particles ($\dr\to 0$). It is then easy to find that $\lambda^*\leq\mu^*$ if $\alpha\leq (4-\dt)/(7\dt-4)$ (i.e., $\alpha<\frac{1}{5}=0.2$ and $\alpha<\frac{1}{17}\simeq 0.06$ for HD and HS, respectively) when the fourth-degree cumulant $a_2$ of $f^{(0)}$ is neglected. Paradoxically, if the role of $a_2$ is introduced in a standard way \cite{BDKS98,BC01}, the interval of absolute instability grows to $\alpha\leq 0.333$ (HD) and $\alpha\leq 0.175$ (HS). However, if the cumulant $a_2$ is taken into account in a more consistent manner \cite{GSM07}, then $\lambda^*>\mu^*$ for all $\alpha$, both for HD and HS. To make things even more complicated, it is known that the cumulant expansion of $f^{(0)}$ for smooth particles breaks down if $\alpha$ is small \cite{BP06,*BP06b}.

The situation is much more delicate in the case of rough particles. First, instead of a single fourth-degree cumulant of $f^{(0)}$, there are three (HD) or four (HS) independent fourth-degree cumulants \cite{paperIII}. Second, those cumulants have been reported for HS \cite{BPKZ07,SKS11,VSK14,VSK14b,VS15}, but not for HD. And third, the known cumulants for HS can take rather large values \cite{VSK14,VSK14b}, except for small inelasticity, and this effect is expected to become even more dramatic for HD \cite{paperIII}.

Considering all of this, the prediction of a region of absolute instability  in the HD case must be taken with much caution. In any case, one can conclude that the HD gas typically develops clustering instabilities with much larger reduced wave numbers than the HS gas if the values of $\alpha$ and $\beta$ belong to the regions signaled in Figs.\ \ref{fig:Locus} and \ref{fig:kpara}.

 \begin{table}
   \caption{Values of the main parameters of the systems analyzed by event-driven MD simulations.}\label{table1}
\begin{ruledtabular}
\begin{tabular}{cccccccc}
System&$N$&$n\sigma^2$&$L/\sigma$&$k$&$M$&$\Lcell/\sigma$&$\langle\Ncell\rangle$\\
\hline
A&$1\,600$&$0.005$&$565.7$&$1.253$&$625$&$22.63$&$2.56$\\
B&$1\,600$&$0.010$&$400.0$&$0.886$&$625$&$16.00$&$2.56$\\
C&$6\,400$&$0.005$&$1\,131.4$&$0.627$&$2\,500$&$22.63$&$2.56$\\
     \end{tabular}
 \end{ruledtabular}
 \end{table}

\begin{figure}
    \includegraphics[width=\columnwidth]{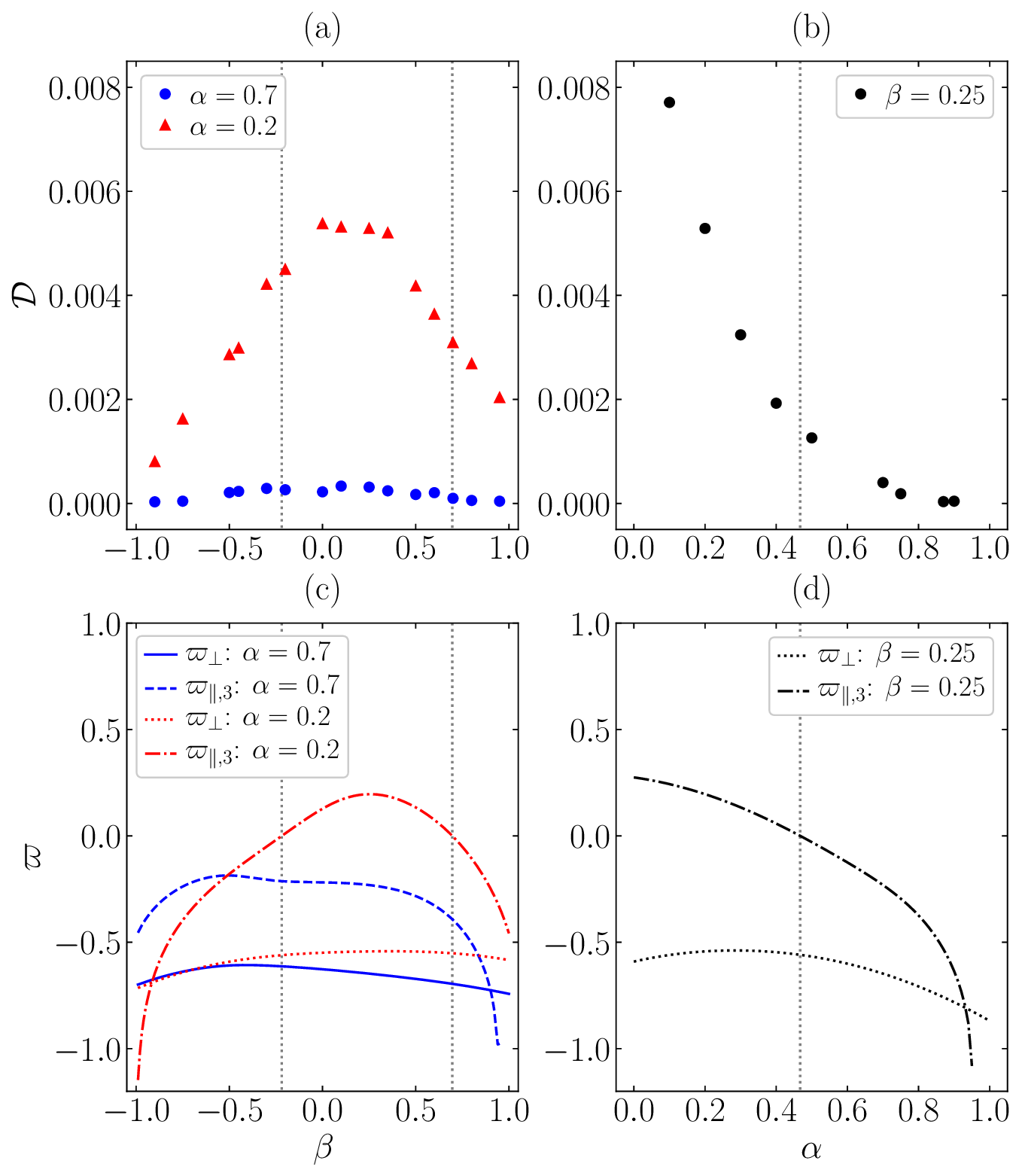}
        \caption{(a) Plot of the MD simulation values of the KLD $\mathcal{D}$ vs $\beta$ at $\alpha=0.2$ and $\alpha=0.7$ for  system A (see Table \ref{table1}). (b) Plot of the MD simulation values of the KLD $\mathcal{D}$ vs $\alpha$ at $\beta=0.25$ for the same system. (c) Theoretical eigenvalues $\varpi_\perp$ (transverse shear mode) and $\varpi_{\|,3}$ (longitudinal heat mode) vs $\beta$ at $\alpha=0.2$ and $\alpha=0.7$ for a reduced wave number $k= 1.253$. (d) Theoretical eigenvalues $\varpi_\perp$ (transverse shear mode) and $\varpi_{\|,3}$ (longitudinal heat mode) vs $\alpha$ at $\beta=0.25$  for a reduced wave number $k= 1.253$. The vertical dotted lines denote the borders of the regions where, according to theory, the HCS of the system is unstable ($\varpi_{\|,3}>0$) at (a, c) $\alpha=0.2$ and (b, d) $\beta=0.25$.}
    \label{fig:KLD}
\end{figure}

\begin{figure}
    \includegraphics[width= \columnwidth]{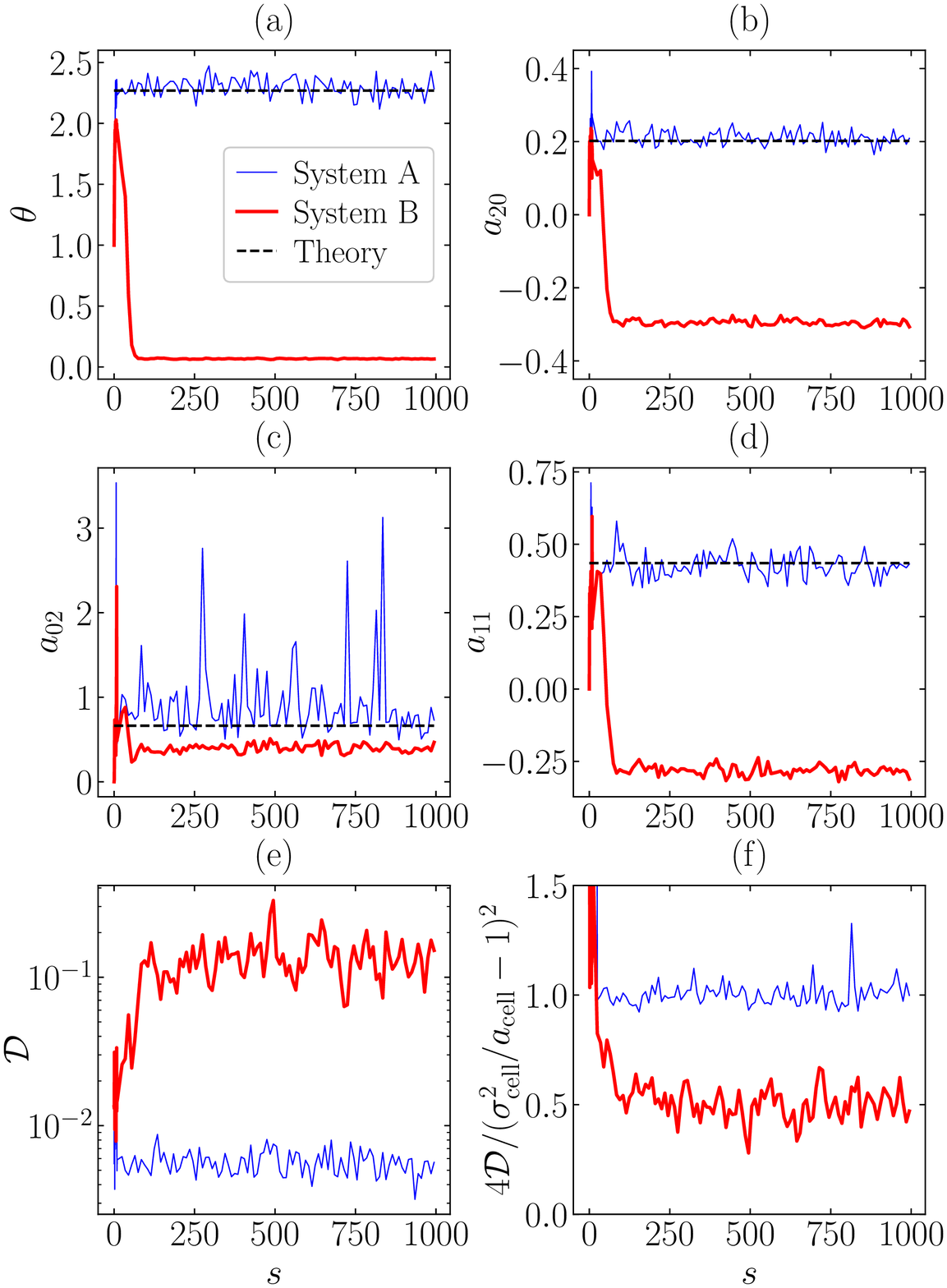}
    \caption{MD temporal evolution of (a) the rotational-to-translational temperature ratio $\theta$, (b) the excess translational velocity kurtosis $a_{20}$, (c) the excess angular velocity kurtosis $a_{02}$, (d) the translational-angular correlation cumulant $a_{11}$, (e) the KLD $\mathcal{D}$, and (f) the ratio $4\mathcal{D}/(\scell^2/\acell-1)^2$ [see Eq.\ \eqref{KLD_approx}]. The (blue) thin and (red) thick lines correspond to  systems A and B, respectively (see Table \ref{table1}),  in both cases with $\alpha=0.2$ and $\beta=0.25$. The horizontal dashed lines in panels (a--d) are theoretical values \cite{paperIII}.}
    \label{fig:comparison}
\end{figure}

\begin{figure}
    \includegraphics[width=\columnwidth]{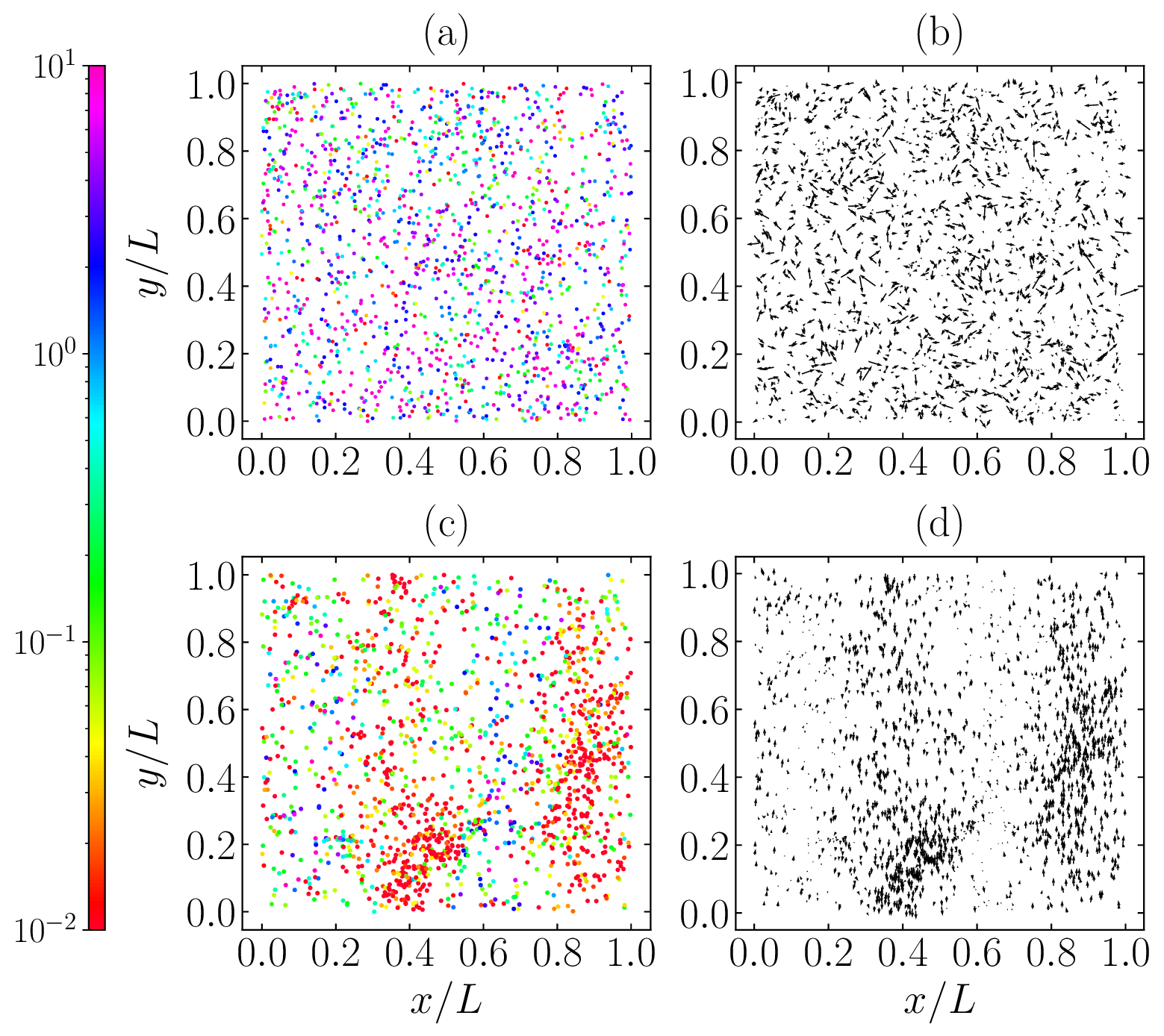}
    \caption{MD snapshots at $s=100$ collisions per particle showing the positions [left panels (a, c)] and translational velocities [right panels (b, d)] in the cases of systems A [top panels (a, b)] and B [bottom panels (c, d)] (see Table \ref{table1}), in both cases with $\alpha=0.2$ and $\beta=0.25$. In the left panels, the color code refers to the ratio between the rotational and the translational kinetic energies of each particle.}
    \label{fig:scatter}
\end{figure}

\begin{figure}
    \includegraphics[width= \columnwidth]{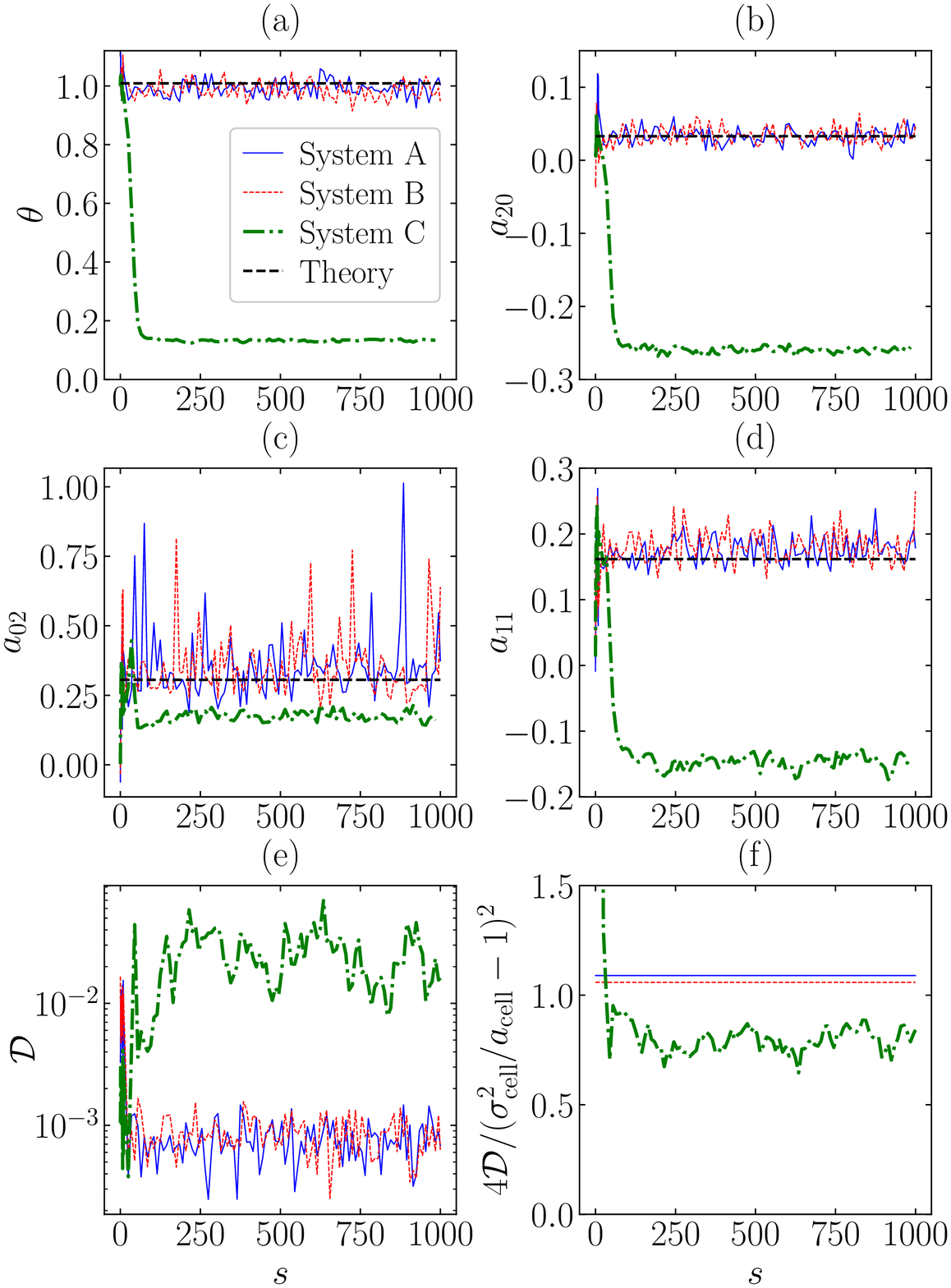}
    \caption{MD temporal evolution of (a) the rotational-to-translational temperature ratio $\theta$, (b) the excess translational velocity kurtosis $a_{20}$, (c) the excess angular velocity kurtosis $a_{02}$, (d) the translational-angular correlation cumulant $a_{11}$, (e) the KLD $\mathcal{D}$, and (f) the ratio $4\mathcal{D}/(\scell^2/\acell-1)^2$ [see Eq.\ \eqref{KLD_approx}]. The (blue) thin solid, the (red) thin dashed, and the (green) thick dash-dotted lines correspond to  systems A, B, and C, respectively (see Table \ref{table1}),  in the three cases with $\alpha=0.7$ and $\beta=0.25$. The horizontal dashed lines in panels (a--d) are theoretical values \cite{paperIII}.}
    \label{fig:comparison_ABC}
\end{figure}

\begin{figure}
    \includegraphics[width=\columnwidth]{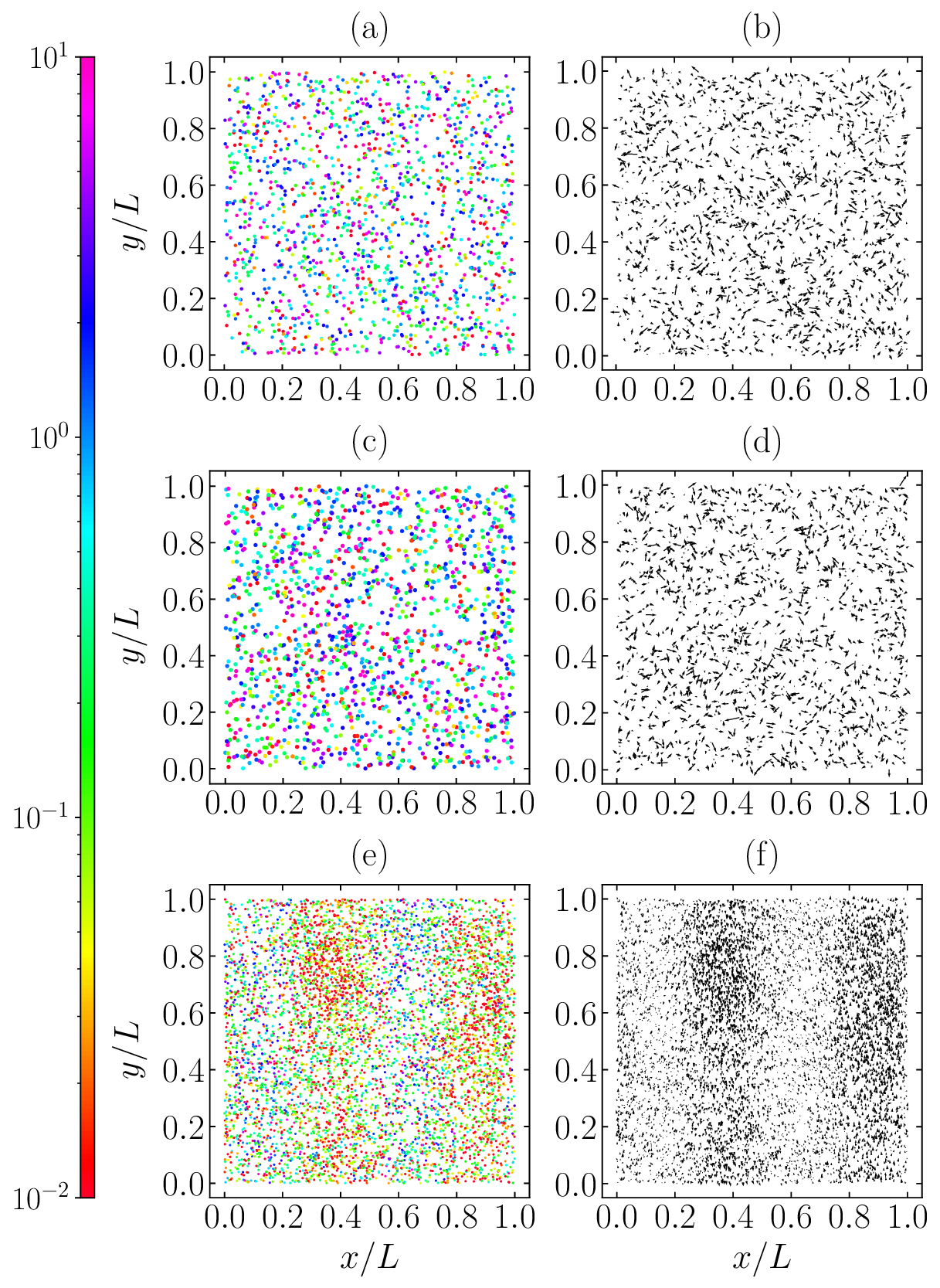}
    \caption{MD snapshots at $s=100$ collisions per particle showing the positions [left panels (a, c, e)] and translational velocities [right panels (b, d, f)] in the cases of systems A [top panels (a, b)],  B [middle panels (c, d)], and C [bottom panels (e, f)] (see Table \ref{table1}), in the three cases with $\alpha=0.7$ and $\beta=0.25$. In the left panels, the color code refers to the ratio between the rotational and the translational kinetic energies of each particle.}
    \label{fig:scatter_ABC}
\end{figure}

\section{Molecular dynamics simulations for inelastic and rough hard disks}\label{sec:5}

Although the main aim of this paper is theoretical, we present in this section  event-driven MD results for freely cooling HD gases to check the stability of the HCS.

We have considered systems characterized by a certain number  $N$ of uniform disks ($\kappa=\frac{1}{2}$) and a certain reduced number density $n\sigma^2$. The particles were enclosed in a square box of side length $L/\sigma=\sqrt{N/n\sigma^2}$ and periodic boundary conditions were applied. Since the largest wavelength of a perturbation is $L$, the smallest (reduced) wave number is $k=2\sqrt{\pi}/n\sigma L=2\sqrt{\pi /N n\sigma^2}$. For each choice of $\alpha$ and $\beta$, the system was allowed to evolve for $s=1\,000$ average number of collisions per particle and data were extracted every $0.5$  collisions per particle.
To avoid dealing with extremely low temperatures and velocities after a large number of collisions per particle, thus compromising the accuracy of the simulation data, a velocity rescaling \cite{L01} was performed every $0.5$ collisions per particle. Moreover, inelastic collapse was prevented by switching to elastic collisions whenever two successive collisions involved the same pair in a very short period of time \cite{LM98}.

The appearance of clustering instabilities in simulations is usually identified by means of visual snapshots \cite{MDHEH13}. However, this might be difficult if $N$ is not  large emough, as happens in a dilute gas. It is then very convenient to monitor the degree of spatial homogeneity of the gas by means of a single quantity that oversees the whole system. To this end, we propose here the (discrete and coarse-grained) KLD  \cite{KL51,K78} of the spatial distribution  of particles in the box with respect to a reference homogeneous distribution as a {control parameter} to detect clustering inhomogeneities. Although the KLD is not actually a metric function, it somehow measures the \emph{distance} (divergence) of a distribution with respect to a reference one as the amount of information lost when the reference model distribution is used to approximate the true distribution. The KLD has been used  to measure inhomogeneities in other physical contexts \cite{HBM04,AC12}. More recently, Shannon's entropy \cite{S48} (which is related to the KLD but with a constant reference distribution)   has been used to study clustering in granular dynamics experiments \cite{SMKSKJDTW21}.

To construct the coarse-grained  KLD,  the HD simulation box is split into   $M\gg 1$ square cells of side length $L_{\mathrm{cell}}$ and area $L_{\mathrm{cell}}^2$. Let us denote by $N_{\mathrm{cell}}=0,1,\ldots,N$ the number of disks inside a given cell. The fraction of cells having exactly $N_{\mathrm{cell}}$ particles  will be denoted as $p(N_{\mathrm{cell}})$; equivalently, this is the probability that a cell chosen at random has $N_{\mathrm{cell}}$ particles. Obviously, the average number of particles per cell is $\acell\equiv\langle{N}_{\mathrm{cell}}\rangle=N/M=n L_{\text{cell}}^2$. A relevant quantity is the variance  $\scell^2=\langle{N}_{\mathrm{cell}}^2\rangle-\acell^2$, measuring fluctuations around the average number.
Now, we define the KLD  as
\BEQ
\label{KLD}
    \mathcal{D} = \sum_{N_{\text{cell}}=0}^{N} p(N_{\text{cell}})\ln\frac{p(N_{\text{cell}})}{p_{\text{ref}}(N_{\text{cell}})},
\EEQ
where $p_{\text{ref}}(N_{\text{cell}})$ is a \emph{reference}  distribution modeling a spatially uniform system. Here we choose such a distribution as that of a system of totally uncorrelated point particles. If we randomly ``shoot'' a particle to the simulation box, then the probability that it hits a given cell is $M^{-1}$. Thus, in the reference model,  the probability that $N_{\text{cell}}$ particles have hit the cell after $N$ shootings is given by the binomial distribution
\BEQ
\label{eq:binom}
    p_{\text{ref}}(N_{\text{cell}}) = M^{-N}\binom{N}{N_{\text{cell}}}  \left(M-1\right)^{N-N_{\text{cell}}}.
\EEQ
In this reference model, $\sigma^2_{\text{cell,ref}}=\acell\left(1-M^{-1}\right)$. Taking into account that $M\gg 1$ (so that $\acell\ll N$), it is possible to approximate the binomial distribution by the Poisson one, $p_{\text{ref}}(\Ncell)\simeq e^{-\acell}\acell^{\Ncell}/\Ncell!$. In that case, $\langle\Ncell^2\rangle_{\text{ref}}\simeq \acell(1+\acell)$, $\langle\Ncell^3\rangle_{\text{ref}}\simeq \acell(1+3\acell+\acell^2)$, and $\langle\Ncell^4\rangle_{\text{ref}}\simeq \acell(1+7\acell+6\acell^2+\acell^3)$.

While the KLD defined by Eq.\ \eqref{KLD} compares the distributions $p(\Ncell)$ and $p_{\text{ref}}(\Ncell)$ for all values of $\Ncell$, a simple relationship between $\mathcal{D}$ and the variance difference $\delta\scell^2\equiv \scell^2-\sigma_{\text{cell,ref}}^2$ can be established if $p(\Ncell)\approx p_{\text{ref}}(\Ncell)$, so we can write
\begin{subequations}
\beq
\label{papprox}
p(\Ncell)\simeq p_{\text{ref}}(\Ncell)\left[1+\frac{\delta\scell^2}{2}S_2(\Ncell)\right],
\eeq
\beq
S_2(\Ncell)=1-\frac{1+2\acell}{\acell^2}\Ncell+\frac{\Ncell^2}{\acell^2}.
\eeq
\end{subequations}
Note that $\langle S_2(\Ncell)\rangle_{\text{ref}}=\langle \Ncell S_2(\Ncell)\rangle_{\text{ref}}=0$ and $\langle \Ncell^2 S_2(\Ncell)\rangle_{\text{ref}}=2$ within the Poisson approximation.
By inserting Eq.\ \eqref{papprox} into Eq.\ \eqref{KLD} and expanding up to second order in $\delta\scell^2$, one gets
\beq
\label{KLD_approx}
\mathcal{D}\simeq \frac{\left(\delta\scell^2\right)^2}{8}\langle [ S_2(\Ncell)]^2\rangle_{\text{ref}}=\frac{1}{4}\left(\frac{\scell^2}{\acell}-1\right)^2.
\eeq
It is important to bear in mind that the reference model neglects excluded-volume effects and nonequilibrium spatial correlations. Therefore, it is possible to have $\mathcal{D}\neq 0$ and $\scell^2\neq \acell\left(1-M^{-1}\right)\simeq \acell$, even if the system remains homogeneous. However,  significant nonzero values of $\mathcal{D}$ and/or  $\scell^2/\acell-1$ are expected to be indicators of spontaneous heterogeneities in the spatial particle distributions.

In most of our simulations, we have chosen the system identified with the label A in Table \ref{table1}. After an aging stage of $s=500$ collisions per particle, the coarse-grained spatial distribution $p(\Ncell)$ was obtained by averaging the histograms corresponding to the population of the $M$ cells from $s=500$ to $s=1\,000$. The KLD was then evaluated from Eqs.\ \eqref{KLD} and \eqref{eq:binom}.
Figures \ref{fig:KLD}(a) and  \ref{fig:KLD}(b) show the dependence of the computed KLD versus $\beta$ (at fixed $\alpha=0.2$ and $\alpha=0.7$) and versus $\alpha$ (at fixed $\beta=0.25$), respectively.
The behavior of the theoretical eigenvalues $\varpi_\perp$ and $\varpi_{\|,3}$ for the value of the wave number corresponding to system A ($k= 1.253$) are shown in Figs.\  \ref{fig:KLD}(c) and  \ref{fig:KLD}(d), respectively. We observe that the MD values of the KLD and the theoretical values of the eigenvalue $\varpi_{\|,3}$ are rather correlated: in general, the larger $\varpi_{\|,3}$ the larger $\mathcal{D}$. The relevant point here is that theory predicts that the system becomes unstable if $\alpha=0.2$ in the interval $-0.217<\beta<0.695$ and if $\beta=0.25$ for $\alpha<0.466$. However, the MD data for $\mathcal{D}$ do not seem to experience a big increase in those cases, thus casting doubts about the true instability of perturbations with $k=1.253$ if $\alpha=0.2$.

To clarify the situation, we have selected the coefficients of restitution $\alpha=0.2$ and $\beta=0.25$, and performed additional simulations for system B (see Table \ref{table1}), in which the associated wave number is $k= 0.886$.
Figure \ref{fig:comparison} shows the temporal evolution of some relevant quantities for both systems (A and B). The considered quantities are (a) the rotational-to translational temperature ratio $\theta=3/\tau_t-2$, (b) the excess  translational velocity kurtosis $a_{20}\equiv \langle V^4\rangle/2\langle V^2\rangle^2-1$, (c) the excess angular velocity kurtosis $a_{02}\equiv \langle \omega^4\rangle/3\langle \omega^2\rangle^2-1$, (d) the translational-angular correlation cumulant $a_{11}\equiv \langle V^2\omega^2\rangle/\langle V^2\rangle\langle \omega^2\rangle-1$, (e) the KLD $\mathcal{D}$, and (f) the ratio $4\mathcal{D}/(\scell^2/\acell-1)^2$.
In a first stage (lasting about $10$ collisions per particle) we have observed that both systems evolve in an analogous way. However, as clearly seen from Fig.\ \ref{fig:comparison}, their evolutions depart from each other in later stages. System B evolves to a state where (a) almost all the kinetic energy is concentrated on the translational degrees of freedom ($\theta\ll 1$), (b) the distribution of translational velocities is strongly platykurtic ($a_{20}<0$), (c) the distribution of angular velocities is much less leptokurtic ($a_{02}>0$) than in system A, (d) the translational velocities are negatively correlated with the angular ones ($a_{11}<0$), (e) the KLD takes  values more than an order of magnitude higher ($\mathcal{D}\sim 10^{-1}$) than in system A, and (f) the estimate given by \eqref{KLD_approx} is much less accurate than in system A. Moreover, the simulation data in the case of system A agree very well with HCS theoretical estimates for $\theta$, $a_{20}$, $a_{02}$, and $a_{11}$ \cite{paperIII}, in sharp contrast to system B.

Figure \ref{fig:comparison} is supplemented by Fig.\ \ref{fig:scatter}, which presents snapshots (at $s=100$) of systems A and B with $\alpha=0.2$ and $\beta=0.25$ \cite{note_21_07_1}. While system A does not present any visible signature of instability, system B exhibits clusters and vortices. Furthermore, the color code in Figs.\ \ref{fig:scatter}(a) and \ref{fig:scatter}(c) shows that disks in system B have typically less rotational energy than translational energy, in contrast to what happens in system A. The loss of rotational energy (relative to the translational one) in system B is stronger in the particles belonging to the clusters, which are also those participating in the vortices and moving with a higher translational velocity.

Therefore, from Figs.\ \ref{fig:comparison} and \ref{fig:scatter} we can conclude that a dilute HD gas with coefficients of restitution $\alpha=0.2$ and $\beta=0.25$ is stable against perturbations of (reduced) wave number $k=1.253$ (system A), while it is unstable against perturbations of (reduced) wave number $k=0.886$ (system B). Thus, the true critical wave number $k_c$ for $\alpha=0.2$ and $\beta=0.25$ must be $0.89<k_c<1.25$. In contrast, in our approximation we obtain $k_\perp=0.822$ but $k_{\|}\to\infty$. As a consequence, a more accurate theoretical treatment of very inelastic particles ($\alpha=0.2$) demands for the inclusion of velocity cumulants in the description.

Let us consider now the case of less inelastic particles, namely $\alpha=0.7$, but still with $\beta=0.25$. In such a case, the theoretical  wave numbers are $k_\perp=0.721$ and $k_\|=0.626$, so that  the clustering instability is preempted by the vortex one and the theoretical critical wave number is $k_c=0.721$. Systems A ($k=1.253$) and B ($k=0.886$) are expected to be stable if  $(\alpha,\beta)=(0.7,0.25)$, despite the fact that  Figs.\ \ref{fig:comparison} and \ref{fig:scatter} showed the instability of system B at  $(\alpha,\beta)=(0.2,0.25)$. To complement the picture, we have also considered the point $(\alpha,\beta)=(0.7,0.25)$ for a third system C (see Table \ref{table1}) for which $k=0.627$; since $k<k_c$, system C is expected to be unstable. The simulation results are displayed in Figs.\ \ref{fig:comparison_ABC} and \ref{fig:scatter_ABC}, which confirm that systems A and B are stable, while system C is unstable \cite{note_21_07_1}.
Note that in Fig.\ \ref{fig:comparison_ABC}(f), due to the low signal-to-noise ratio of the evolution curves of both $\mathcal{D}$ and $(\scell^2/\acell-1)^2/4$ in systems A and B, only the steady-state ratio $4\mathcal{D}/(\scell^2/\acell-1)^2$ is shown in the case of  those systems.

Thus, according to our MD simulations, the true critical wave number for $(\alpha,\beta)=(0.7,0.25)$ lies in the interval $0.63<k_c<0.89$, in close agreement with the theoretical prediction $k_c=0.721$. Moreover, $k< k_\perp$ and, as can be observed from Figs.\ \ref{fig:comparison_ABC} and \ref{fig:scatter_ABC},  clustering is indeed present, which means that the theoretical prediction is pretty reliable for this moderately inelastic case.

\section{Concluding remarks}\label{sec:6}

In this work, we have carried out a detailed linear stability analysis of the HCS  of a dilute gas of inelastic and rough HD or HS within a common framework, thus extending previous HS results  \cite{GSK18} to the case of HD gases.  First, the NSF equations have been linearized around the HCS solution by a formally exact  analysis. Next, the final results have been obtained by the introduction of the approximate expressions of the transport coefficients derived in the companion paper I \cite{paperI}, which are nonlinear functions of the coefficients of normal ($\alpha$) and tangential ($\beta$) restitution, the reduced moment of inertia ($\kappa$), and the numbers of degrees of freedom ($\dt$ and $\dr$).

As happens with rough HS \cite{GSK18} and the case of $\dt$-dimensional smooth particles \cite{BDKS98,BC01,GSM07}, there are two longitudinal (sound) modes that are always stable, whereas the third longitudinal (heat) mode and the $(\dt-1)$-fold transverse (shear) modes become unstable for long enough wavelengths. The heat mode is associated with cluster instabilities, while the shear modes are related to  vortex formation. This analysis has allowed us to determine the critical length $L_c$, such that systems with a size $L>L_c$ are unstable under linear perturbations. The outcome highlights that, in general, two-dimensional HD systems become unstable for smaller reduced wavelengths than their three-dimensional HS counterparts. Additionally, the dual role of roughness, according to which small and large levels of roughness make the system less unstable than the frictionless system, previously observed in the HS geometry \cite{MDHEH13,GSK18}, still holds in the HD case. Moreover, we have established that the region in the parameter space where  cluster instabilities dominate against vortices (i.e.,  $k_{\parallel}>k_{\perp}$) is generally larger for HD than for HS.

The most surprising consequence of our analysis is the appearance of a  region of absolute instability, where  the critical longitudinal wave number diverges ($k_{\parallel}\to\infty$ or, equivalently, $L_c\rightarrow 0$). The boundary of this region is defined by the condition $\lambda^{*}=\mu^{*}$, which, while  residually present in HS systems, is especially relevant in the HD case (see Fig.\ \ref{fig:Locus}). In fact, the HS region of absolute instability vanishes if $\kappa>0.277$ (what includes the case of a uniform mass distribution, $\kappa=\frac{2}{5}$) but  it always emerges in the  HD case, regardless of the value of $\kappa$.

The absolute instability zone for HD is a very peculiar prediction, and one must be wary of it. First of all, we have established that this region materializes for very inelastic systems (at least $\alpha<0.426$ if $\kappa=0.302$ and $\alpha<0.392$ if $\kappa=\frac{1}{2}$). Even for the smooth case, one can face a similar issue in standard approximations \cite{BDKS98,BC01}, which disappears if a more consistent approach is employed \cite{GSM07}. In addition, it is known for HS that velocity cumulants in the HCS may play an important role \cite{SKS11,VSK14,VSK14b}, its effect being even more noticeable for HD \cite{paperIII}. Therefore, to study whether the absolute instability phenomenon actually exists or  is an artifact of the performed approximations, we have carried out event-driven MD simulations which address this question.

To deal with the problem, small system sizes must be tested in the simulations, which implies a small number of particles in the dilute case. Because of that, we have chosen not to rely only on a visual determination of clustering or vortices  via snapshots. This fact was the clincher to use a coarse-grained KLD  (with a binomial distribution as the reference probability distribution) to monitor the presence of spatial heterogeneities. Moreover, instead of analyzing deviations from Haff's cooling law as indicators of instability \cite{MDCPH11,PDR14}, we have focused on the temporal evolution of quantities (such as the rotational-to-translational temperature ratio $\theta$ and velocity cumulants) that are unaffected by the velocity scaling performed in our simulations.

Two-dimensional MD simulations of HD with a uniform mass distribution ($\kappa=\frac{1}{2}$) were established under three different setups  (A, B, and C), as summarized in Table \ref{table1}.
The solid fraction $\phi=\frac{\pi}{4}n\sigma^2$ of each system is low enough as to expect the Boltzmann description for dilute gases to be applicable. For instance, the Enskog factor is $1.006$ (systems A and C) and $1.012$ (system B). The reliability of the Boltzmann equation is also supported by the good agreement between theory and simulations observed for the temperature ratio $\theta$ and the cumulants $a_{20}$, $a_{02}$, and $a_{11}$ in Fig.\ \ref{fig:comparison} for system A and in Fig.\ \ref{fig:comparison_ABC} for systems A and B.

The high-inelasticity point $(\alpha,\beta)=(0.2,0.25)$ lies inside the theoretical region of absolute instability. However, according to Figs.\ \ref{fig:comparison} and \ref{fig:scatter}, although system B ($k\simeq 0.89$) is indeed unstable, system A ($k\simeq 1.25$) is not.
Thus, the (reduced) critical wave number at $(\alpha,\beta)=(0.2,0.25)$ does not diverge but is bounded as $0.89<k_c<1.25$; this critical value is anyway relatively high, as compared with HS values or with values in other regions of the HD parameter space (see Figs.\ \ref{fig:kperp} and \ref{fig:kpara}).
The picture is complemented with the moderate-inelasticity point $(\alpha,\beta)=(0.7,0.25)$, in which case systems A and B are stable, while system C ($k\simeq 0.63$) is not (see Figs.\ \ref{fig:comparison_ABC} and \ref{fig:scatter_ABC}). The determined range $0.63<k_c<0.89$ is now consistent with the theoretical prediction $k_c=0.721$.

It is worth noting that in the cases where our MD simulations indicated instability (system B in Fig.\ \ref{fig:comparison}, system C in Fig.\ \ref{fig:comparison_ABC}), the temperature ratio $\theta$ reached small but nonzero stationary values after a certain number of collisions per particle. This implies a dramatic loss of rotational energy relative to the translational one, which is stronger in the particles involved in cluster and vortex formation. The fact that $\lim_{t\to\infty}\theta(t)\neq 0$ in the unstable regime contrasts with results for moderately dense HD systems reported in Ref.\ \cite{PDR14}, according to which $\theta(t)\sim t^{-0.6}\to 0$.
A possible explanation is that the different cooling power laws observed in Ref.\ \cite{PDR14} may be present in a transient evolution stage, but for a sufficiently large number of collisions per particle both average energies reach a common decay and thus an asymptotic stationary value $\theta\neq 0$ is obtained.

While signaling a region of strong instability, the predicted high-inelasticity region of absolute instability seems to be a consequence  of the neglect of HCS velocity cumulants in the derivation of the NSF transport coefficients carried out in paper I \cite{paperI}. This calls for a more complex and consistent treatment  which we plan to undertake in the near future \cite{paperIII}. We will also carry out a similar work for stochastically driven granular gases, in which case the ansatz of a semi-Maxwellian form for the velocity distribution function of the base reference state is more accurate than in the free cooling situation.

To conclude, we hope this work may encourage further investigation on this topic,  such as better approximations, more computer simulations by both MD and the direct simulation Monte Carlo (DSMC) method, and even experimental tests about the impact of roughness on  the hydrodynamic properties and stability of HD and HS granular gases.

\begin{acknowledgments}
The authors acknowledge financial support from the Grant No.\ PID2020-112936GB-I00/AEI/10.13039/501100011033 and from the Junta de Extremadura (Spain) through Grants No.\ IB20079 and No.\ GR18079, all of them partially financed by Fondo Europeo de Desarrollo Regional funds. A.M. is grateful to the Spanish Ministerio de Ciencia, Innovaci\'on y Universidades for support from a predoctoral fellowship Grant No.\ FPU2018-3503.
\end{acknowledgments}

%\bibliographystyle{apsrev}

%\bibliography{C:/AA_D/Dropbox/Mis_Dropcumentos/bib_files/Granular.bib}

\begin{thebibliography}{38}%
\makeatletter
\providecommand \@ifxundefined [1]{%
 \@ifx{#1\undefined}
}%
\providecommand \@ifnum [1]{%
 \ifnum #1\expandafter \@firstoftwo
 \else \expandafter \@secondoftwo
 \fi
}%
\providecommand \@ifx [1]{%
 \ifx #1\expandafter \@firstoftwo
 \else \expandafter \@secondoftwo
 \fi
}%
\providecommand \natexlab [1]{#1}%
\providecommand \enquote  [1]{``#1''}%
\providecommand \bibnamefont  [1]{#1}%
\providecommand \bibfnamefont [1]{#1}%
\providecommand \citenamefont [1]{#1}%
\providecommand \href@noop [0]{\@secondoftwo}%
\providecommand \href [0]{\begingroup \@sanitize@url \@href}%
\providecommand \@href[1]{\@@startlink{#1}\@@href}%
\providecommand \@@href[1]{\endgroup#1\@@endlink}%
\providecommand \@sanitize@url [0]{\catcode `\\12\catcode `\$12\catcode
  `\&12\catcode `\#12\catcode `\^12\catcode `\_12\catcode `\%12\relax}%
\providecommand \@@startlink[1]{}%
\providecommand \@@endlink[0]{}%
\providecommand \url  [0]{\begingroup\@sanitize@url \@url }%
\providecommand \@url [1]{\endgroup\@href {#1}{\urlprefix }}%
\providecommand \urlprefix  [0]{URL }%
\providecommand \Eprint [0]{\href }%
\providecommand \doibase [0]{https://doi.org/}%
\providecommand \selectlanguage [0]{\@gobble}%
\providecommand \bibinfo  [0]{\@secondoftwo}%
\providecommand \bibfield  [0]{\@secondoftwo}%
\providecommand \translation [1]{[#1]}%
\providecommand \BibitemOpen [0]{}%
\providecommand \bibitemStop [0]{}%
\providecommand \bibitemNoStop [0]{.\EOS\space}%
\providecommand \EOS [0]{\spacefactor3000\relax}%
\providecommand \BibitemShut  [1]{\csname bibitem#1\endcsname}%
\let\auto@bib@innerbib\@empty
%</preamble>
\bibitem [{\citenamefont {Goldhirsch}\ and\ \citenamefont
  {Zanetti}(1993)}]{GZ93}%
  \BibitemOpen
  \bibfield  {author} {\bibinfo {author} {\bibfnamefont {I.}~\bibnamefont
  {Goldhirsch}}\ and\ \bibinfo {author} {\bibfnamefont {G.}~\bibnamefont
  {Zanetti}},\ }\bibfield  {title} {\bibinfo {title} {Clustering instability in
  dissipative gases},\ }\href {https://doi.org/10.1103/PhysRevLett.70.1619}
  {\bibfield  {journal} {\bibinfo  {journal} {Phys. Rev. Lett.}\ }\textbf
  {\bibinfo {volume} {70}},\ \bibinfo {pages} {1619} (\bibinfo {year}
  {1993})}\BibitemShut {NoStop}%
\bibitem [{\citenamefont {McNamara}(1993)}]{M93b}%
  \BibitemOpen
  \bibfield  {author} {\bibinfo {author} {\bibfnamefont {S.}~\bibnamefont
  {McNamara}},\ }\bibfield  {title} {\bibinfo {title} {Hydrodynamic modes of a
  uniform granular medium},\ }\href {https://doi.org/10.1063/1.858716}
  {\bibfield  {journal} {\bibinfo  {journal} {Phys. Fluids A}\ }\textbf
  {\bibinfo {volume} {5}},\ \bibinfo {pages} {3056} (\bibinfo {year}
  {1993})}\BibitemShut {NoStop}%
\bibitem [{\citenamefont {McNamara}\ and\ \citenamefont {Young}(1994)}]{MY94}%
  \BibitemOpen
  \bibfield  {author} {\bibinfo {author} {\bibfnamefont {S.}~\bibnamefont
  {McNamara}}\ and\ \bibinfo {author} {\bibfnamefont {W.~R.}\ \bibnamefont
  {Young}},\ }\bibfield  {title} {\bibinfo {title} {Inelastic collapse in two
  dimensions},\ }\href {https://doi.org/10.1103/PhysRevE.50.R28} {\bibfield
  {journal} {\bibinfo  {journal} {Phys. Rev. E}\ }\textbf {\bibinfo {volume}
  {50}},\ \bibinfo {pages} {R28} (\bibinfo {year} {1994})}\BibitemShut
  {NoStop}%
\bibitem [{\citenamefont {McNamara}\ and\ \citenamefont {Young}(1996)}]{MY96}%
  \BibitemOpen
  \bibfield  {author} {\bibinfo {author} {\bibfnamefont {S.}~\bibnamefont
  {McNamara}}\ and\ \bibinfo {author} {\bibfnamefont {W.~R.}\ \bibnamefont
  {Young}},\ }\bibfield  {title} {\bibinfo {title} {Dynamics of a freely
  evolving, two-dimensional granular medium},\ }\href
  {https://doi.org/10.1103/PhysRevE.53.5089} {\bibfield  {journal} {\bibinfo
  {journal} {Phys. Rev. E}\ }\textbf {\bibinfo {volume} {53}},\ \bibinfo
  {pages} {5089} (\bibinfo {year} {1996})}\BibitemShut {NoStop}%
\bibitem [{\citenamefont {Brey}\ \emph {et~al.}(1998)\citenamefont {Brey},
  \citenamefont {Dufty}, \citenamefont {Kim},\ and\ \citenamefont
  {Santos}}]{BDKS98}%
  \BibitemOpen
  \bibfield  {author} {\bibinfo {author} {\bibfnamefont {J.~J.}\ \bibnamefont
  {Brey}}, \bibinfo {author} {\bibfnamefont {J.~W.}\ \bibnamefont {Dufty}},
  \bibinfo {author} {\bibfnamefont {C.~S.}\ \bibnamefont {Kim}},\ and\ \bibinfo
  {author} {\bibfnamefont {A.}~\bibnamefont {Santos}},\ }\bibfield  {title}
  {\bibinfo {title} {Hydrodynamics for granular flow at low density},\ }\href
  {https://doi.org/10.1103/PhysRevE.58.4638} {\bibfield  {journal} {\bibinfo
  {journal} {Phys. Rev. E}\ }\textbf {\bibinfo {volume} {58}},\ \bibinfo
  {pages} {4638} (\bibinfo {year} {1998})}\BibitemShut {NoStop}%
\bibitem [{\citenamefont {Luding}\ and\ \citenamefont {Herrmann}(1999)}]{LH99}%
  \BibitemOpen
  \bibfield  {author} {\bibinfo {author} {\bibfnamefont {S.}~\bibnamefont
  {Luding}}\ and\ \bibinfo {author} {\bibfnamefont {H.~J.}\ \bibnamefont
  {Herrmann}},\ }\bibfield  {title} {\bibinfo {title} {Cluster-growth in freely
  cooling granular media},\ }\href {https://doi.org/10.1063/1.166441}
  {\bibfield  {journal} {\bibinfo  {journal} {Chaos}\ }\textbf {\bibinfo
  {volume} {9}},\ \bibinfo {pages} {673} (\bibinfo {year} {1999})}\BibitemShut
  {NoStop}%
\bibitem [{\citenamefont {Fullmer}\ and\ \citenamefont {Hrenya}(2017)}]{FH17}%
  \BibitemOpen
  \bibfield  {author} {\bibinfo {author} {\bibfnamefont {W.~D.}\ \bibnamefont
  {Fullmer}}\ and\ \bibinfo {author} {\bibfnamefont {C.~M.}\ \bibnamefont
  {Hrenya}},\ }\bibfield  {title} {\bibinfo {title} {The clustering instability
  in rapid granular and gas-solid flows},\ }\href
  {https://doi.org/10.1146/annurev-fluid-010816-060028} {\bibfield  {journal}
  {\bibinfo  {journal} {Annu. Rev. Fluid Mech.}\ }\textbf {\bibinfo {volume}
  {49}},\ \bibinfo {pages} {485} (\bibinfo {year} {2017})}\BibitemShut
  {NoStop}%
\bibitem [{\citenamefont {Garz\'o}(2019)}]{G19}%
  \BibitemOpen
  \bibfield  {author} {\bibinfo {author} {\bibfnamefont {V.}~\bibnamefont
  {Garz\'o}},\ }\href@noop {} {\emph {\bibinfo {title} {Granular Gaseous Flows.
  A Kinetic Theory Approach to Granular Gaseous Flows}}}\ (\bibinfo
  {publisher} {Springer Nature},\ \bibinfo {address} {Switzerland},\ \bibinfo
  {year} {2019})\BibitemShut {NoStop}%
\bibitem [{\citenamefont {Mitrano}\ \emph {et~al.}(2013)\citenamefont
  {Mitrano}, \citenamefont {Dahl}, \citenamefont {Hilger}, \citenamefont
  {Ewasko},\ and\ \citenamefont {Hrenya}}]{MDHEH13}%
  \BibitemOpen
  \bibfield  {author} {\bibinfo {author} {\bibfnamefont {P.~P.}\ \bibnamefont
  {Mitrano}}, \bibinfo {author} {\bibfnamefont {S.~R.}\ \bibnamefont {Dahl}},
  \bibinfo {author} {\bibfnamefont {A.~M.}\ \bibnamefont {Hilger}}, \bibinfo
  {author} {\bibfnamefont {C.~J.}\ \bibnamefont {Ewasko}},\ and\ \bibinfo
  {author} {\bibfnamefont {C.~M.}\ \bibnamefont {Hrenya}},\ }\bibfield  {title}
  {\bibinfo {title} {Dual role of friction in granular flows: attenuation
  versus enhancement of instabilities},\ }\href
  {https://doi.org/10.1017/jfm.2013.328} {\bibfield  {journal} {\bibinfo
  {journal} {J. Fluid Mech.}\ }\textbf {\bibinfo {volume} {729}},\ \bibinfo
  {pages} {484} (\bibinfo {year} {2013})}\BibitemShut {NoStop}%
\bibitem [{\citenamefont {Garz\'o}\ \emph {et~al.}(2018)\citenamefont
  {Garz\'o}, \citenamefont {Santos},\ and\ \citenamefont {Kremer}}]{GSK18}%
  \BibitemOpen
  \bibfield  {author} {\bibinfo {author} {\bibfnamefont {V.}~\bibnamefont
  {Garz\'o}}, \bibinfo {author} {\bibfnamefont {A.}~\bibnamefont {Santos}},\
  and\ \bibinfo {author} {\bibfnamefont {G.~M.}\ \bibnamefont {Kremer}},\
  }\bibfield  {title} {\bibinfo {title} {Impact of roughness on the instability
  of a free-cooling granular gas},\ }\href
  {https://doi.org/10.1103/PhysRevE.97.052901} {\bibfield  {journal} {\bibinfo
  {journal} {Phys. Rev. E}\ }\textbf {\bibinfo {volume} {97}},\ \bibinfo
  {pages} {{052}{901}} (\bibinfo {year} {2018})}\BibitemShut {NoStop}%
\bibitem [{\citenamefont {Kremer}(2020)}]{K20}%
  \BibitemOpen
  \bibfield  {author} {\bibinfo {author} {\bibfnamefont {G.~M.}\ \bibnamefont
  {Kremer}},\ }\bibfield  {title} {\bibinfo {title} {Instabilities in a
  self-gravitating granular gas},\ }\href
  {https://doi.org/10.1016/j.physa.2019.123667} {\bibfield  {journal} {\bibinfo
   {journal} {Physica A}\ }\textbf {\bibinfo {volume} {545}},\ \bibinfo {pages}
  {123667} (\bibinfo {year} {2020})}\BibitemShut {NoStop}%
\bibitem [{\citenamefont {Brilliantov}\ \emph {et~al.}(2015)\citenamefont
  {Brilliantov}, \citenamefont {Krapivsky}, \citenamefont {Bodrova},
  \citenamefont {Spahn}, \citenamefont {Hayakawa}, \citenamefont {Stadnichuk},\
  and\ \citenamefont {Schmidt}}]{BKBSHSS15}%
  \BibitemOpen
  \bibfield  {author} {\bibinfo {author} {\bibfnamefont {N.}~\bibnamefont
  {Brilliantov}}, \bibinfo {author} {\bibfnamefont {P.~L.}\ \bibnamefont
  {Krapivsky}}, \bibinfo {author} {\bibfnamefont {A.}~\bibnamefont {Bodrova}},
  \bibinfo {author} {\bibfnamefont {F.}~\bibnamefont {Spahn}}, \bibinfo
  {author} {\bibfnamefont {H.}~\bibnamefont {Hayakawa}}, \bibinfo {author}
  {\bibfnamefont {V.}~\bibnamefont {Stadnichuk}},\ and\ \bibinfo {author}
  {\bibfnamefont {J.}~\bibnamefont {Schmidt}},\ }\bibfield  {title} {\bibinfo
  {title} {Size distribution of particles in {S}aturn's rings from aggregation
  and fragmentation},\ }\href {https://doi.org/10.1073/pnas.1503957112}
  {\bibfield  {journal} {\bibinfo  {journal} {Proc. Natl. Acad. Sci. U. S. A.}\
  }\textbf {\bibinfo {volume} {112}},\ \bibinfo {pages} {9536} (\bibinfo {year}
  {2015})}\BibitemShut {NoStop}%
\bibitem [{\citenamefont {Ballouz}\ \emph {et~al.}(2017)\citenamefont
  {Ballouz}, \citenamefont {Richardson},\ and\ \citenamefont
  {Morishima}}]{BRC17}%
  \BibitemOpen
  \bibfield  {author} {\bibinfo {author} {\bibfnamefont {R.-L.}\ \bibnamefont
  {Ballouz}}, \bibinfo {author} {\bibfnamefont {D.~C.}\ \bibnamefont
  {Richardson}},\ and\ \bibinfo {author} {\bibfnamefont {R.}~\bibnamefont
  {Morishima}},\ }\bibfield  {title} {\bibinfo {title} {Numerical simulations
  of {S}aturn's {B} ring: Granular friction as a mediator between self-gravity
  wakes and viscous overstability},\ }\href
  {https://doi.org/10.3847/1538-3881/aa60be} {\bibfield  {journal} {\bibinfo
  {journal} {Astron. J.}\ }\textbf {\bibinfo {volume} {153}},\ \bibinfo {pages}
  {146} (\bibinfo {year} {2017})}\BibitemShut {NoStop}%
\bibitem [{\citenamefont {Meg\'ias}\ and\ \citenamefont
  {Santos}(2019{\natexlab{a}})}]{MS19}%
  \BibitemOpen
  \bibfield  {author} {\bibinfo {author} {\bibfnamefont {A.}~\bibnamefont
  {Meg\'ias}}\ and\ \bibinfo {author} {\bibfnamefont {A.}~\bibnamefont
  {Santos}},\ }\bibfield  {title} {\bibinfo {title} {Driven and undriven states
  of multicomponent granular gases of inelastic and rough hard disks or
  spheres},\ }\href {https://doi.org/10.1007/s10035-019-0901-y} {\bibfield
  {journal} {\bibinfo  {journal} {Granul. Matter}\ }\textbf {\bibinfo {volume}
  {21}},\ \bibinfo {pages} {49} (\bibinfo {year}
  {2019}{\natexlab{a}})}\BibitemShut {NoStop}%
\bibitem [{\citenamefont {Meg\'ias}\ and\ \citenamefont
  {Santos}(2019{\natexlab{b}})}]{MS19b}%
  \BibitemOpen
  \bibfield  {author} {\bibinfo {author} {\bibfnamefont {A.}~\bibnamefont
  {Meg\'ias}}\ and\ \bibinfo {author} {\bibfnamefont {A.}~\bibnamefont
  {Santos}},\ }\bibfield  {title} {\bibinfo {title} {Energy production rates of
  multicomponent granular gases of rough particles. a unified view of hard-disk
  and hard-sphere systems},\ }\href {https://doi.org/10.1063/1.5119584}
  {\bibfield  {journal} {\bibinfo  {journal} {AIP Conf. Proc.}\ }\textbf
  {\bibinfo {volume} {2132}},\ \bibinfo {pages} {{080}{003}} (\bibinfo {year}
  {2019}{\natexlab{b}})}\BibitemShut {NoStop}%
\bibitem [{\citenamefont {Meg\'ias}\ and\ \citenamefont
  {Santos}(2021{\natexlab{b}})}]{paperI}%
  \BibitemOpen
  \bibfield  {author} {\bibinfo {author} {\bibfnamefont {A.}~\bibnamefont
  {Meg\'ias}}\ and\ \bibinfo {author} {\bibfnamefont {A.}~\bibnamefont
  {Santos}},\ }\bibfield  {title} {\bibinfo {title} {Hydrodynamics of granular gases
  of inelastic and rough hard disks or spheres. {I}. {T}ransport
  coefficients},\ }\href
  {https://doi.org/10.1103/PhysRevE.104.034901} {\bibfield  {journal} {\bibinfo
  {journal} {Phys. Rev. E}\ }\textbf {\bibinfo {volume} {104}},\ \bibinfo
  {pages} {034901} (\bibinfo {year} {2021})}\BibitemShut {NoStop}%  
\bibitem [{\citenamefont {Kullback}\ and\ \citenamefont
  {Leibler}(1951)}]{KL51}%
  \BibitemOpen
  \bibfield  {author} {\bibinfo {author} {\bibfnamefont {S.}~\bibnamefont
  {Kullback}}\ and\ \bibinfo {author} {\bibfnamefont {R.~A.}\ \bibnamefont
  {Leibler}},\ }\bibfield  {title} {\bibinfo {title} {On information and
  sufficiency},\ }\href {https://doi.org/10.1214/aoms/1177729694} {\bibfield
  {journal} {\bibinfo  {journal} {Ann. Math. Statist.}\ }\textbf {\bibinfo
  {volume} {22}},\ \bibinfo {pages} {79} (\bibinfo {year} {1951})}\BibitemShut
  {NoStop}%
\bibitem [{\citenamefont {Kullback}(1978)}]{K78}%
  \BibitemOpen
  \bibfield  {author} {\bibinfo {author} {\bibfnamefont {S.}~\bibnamefont
  {Kullback}},\ }\href@noop {} {\emph {\bibinfo {title} {Information Theory and
  Statistics}}}\ (\bibinfo  {publisher} {Dover},\ \bibinfo {address} {New
  York},\ \bibinfo {year} {1978})\BibitemShut {NoStop}%
\bibitem [{\citenamefont {Pathak}\ \emph {et~al.}(2014)\citenamefont {Pathak},
  \citenamefont {Das},\ and\ \citenamefont {Rajesh}}]{PDR14}%
  \BibitemOpen
  \bibfield  {author} {\bibinfo {author} {\bibfnamefont {S.~N.}\ \bibnamefont
  {Pathak}}, \bibinfo {author} {\bibfnamefont {D.}~\bibnamefont {Das}},\ and\
  \bibinfo {author} {\bibfnamefont {R.}~\bibnamefont {Rajesh}},\ }\bibfield
  {title} {\bibinfo {title} {Inhomogeneous cooling of the rough granular gas in
  two dimensions},\ }\href {https://doi.org/10.1209/0295-5075/107/44001}
  {\bibfield  {journal} {\bibinfo  {journal} {EPL}\ }\textbf {\bibinfo {volume}
  {107}},\ \bibinfo {pages} {44001} (\bibinfo {year} {2014})}\BibitemShut
  {NoStop}%
\bibitem [{\citenamefont {Kremer}\ \emph {et~al.}(2014)\citenamefont {Kremer},
  \citenamefont {Santos},\ and\ \citenamefont {Garz\'o}}]{KSG14}%
  \BibitemOpen
  \bibfield  {author} {\bibinfo {author} {\bibfnamefont {G.~M.}\ \bibnamefont
  {Kremer}}, \bibinfo {author} {\bibfnamefont {A.}~\bibnamefont {Santos}},\
  and\ \bibinfo {author} {\bibfnamefont {V.}~\bibnamefont {Garz\'o}},\
  }\bibfield  {title} {\bibinfo {title} {Transport coefficients of a granular
  gas of inelastic rough hard spheres},\ }\href
  {https://doi.org/10.1103/PhysRevE.90.022205} {\bibfield  {journal} {\bibinfo
  {journal} {Phys. Rev. E}\ }\textbf {\bibinfo {volume} {90}},\ \bibinfo
  {pages} {{022}{205}} (\bibinfo {year} {2014})}\BibitemShut {NoStop}%
\bibitem [{\citenamefont {Brey}\ and\ \citenamefont {Cubero}(2001)}]{BC01}%
  \BibitemOpen
  \bibfield  {author} {\bibinfo {author} {\bibfnamefont {J.~J.}\ \bibnamefont
  {Brey}}\ and\ \bibinfo {author} {\bibfnamefont {D.}~\bibnamefont {Cubero}},\
  }\bibfield  {title} {\bibinfo {title} {Hydrodynamic transport coefficients of
  granular gases},\ }in\ \href@noop {} {\emph {\bibinfo {booktitle} {Granular
  Gases}}},\ \bibinfo {series} {Lectures Notes in Physics}, Vol.\ \bibinfo
  {volume} {564},\ \bibinfo {editor} {edited by\ \bibinfo {editor}
  {\bibfnamefont {T.}~\bibnamefont {P\"oschel}}\ and\ \bibinfo {editor}
  {\bibfnamefont {S.}~\bibnamefont {Luding}}}\ (\bibinfo  {publisher}
  {Springer},\ \bibinfo {address} {Berlin},\ \bibinfo {year} {2001})\ pp.\
  \bibinfo {pages} {59--78}\BibitemShut {NoStop}%
\bibitem [{\citenamefont {Garz\'o}\ \emph {et~al.}(2007)\citenamefont
  {Garz\'o}, \citenamefont {Santos},\ and\ \citenamefont {Montanero}}]{GSM07}%
  \BibitemOpen
  \bibfield  {author} {\bibinfo {author} {\bibfnamefont {V.}~\bibnamefont
  {Garz\'o}}, \bibinfo {author} {\bibfnamefont {A.}~\bibnamefont {Santos}},\
  and\ \bibinfo {author} {\bibfnamefont {J.~M.}\ \bibnamefont {Montanero}},\
  }\bibfield  {title} {\bibinfo {title} {Modified {Sonine} approximation for
  the {Navier--Stokes} transport coefficients of a granular gas},\ }\href
  {https://doi.org/10.1016/j.physa.2006.10.081} {\bibfield  {journal} {\bibinfo
   {journal} {Physica A}\ }\textbf {\bibinfo {volume} {376}},\ \bibinfo {pages}
  {94} (\bibinfo {year} {2007})}\BibitemShut {NoStop}%
\bibitem [{\citenamefont {Brilliantov}\ and\ \citenamefont
  {P\"oschel}(2006{\natexlab{a}})}]{BP06}%
  \BibitemOpen
  \bibfield  {author} {\bibinfo {author} {\bibfnamefont {N.}~\bibnamefont
  {Brilliantov}}\ and\ \bibinfo {author} {\bibfnamefont {T.}~\bibnamefont
  {P\"oschel}},\ }\bibfield  {title} {\bibinfo {title} {Breakdown of the
  {Sonine} expansion for the velocity distribution of granular gases},\ }\href
  {https://doi.org/10.1209/epl/i2005-10555-6} {\bibfield  {journal} {\bibinfo
  {journal} {Europhys. Lett.}\ }\textbf {\bibinfo {volume} {74}},\ \bibinfo
  {pages} {424} (\bibinfo {year} {2006}{\natexlab{a}})}\BibitemShut {NoStop}%
\bibitem [{\citenamefont {Brilliantov}\ and\ \citenamefont
  {P\"oschel}(2006{\natexlab{b}})}]{BP06b}%
  \BibitemOpen
  \bibfield  {author} {\bibinfo {author} {\bibfnamefont {N.}~\bibnamefont
  {Brilliantov}}\ and\ \bibinfo {author} {\bibfnamefont {T.}~\bibnamefont
  {P\"oschel}},\ }\bibfield  {title} {\bibinfo {title} {Erratum: Breakdown of
  the {Sonine} expansion for the velocity distribution of granular gases},\
  }\href {https://doi.org/10.1209/epl/i2006-10099-3} {\bibfield  {journal}
  {\bibinfo  {journal} {Europhys. Lett.}\ }\textbf {\bibinfo {volume} {75}},\
  \bibinfo {pages} {188} (\bibinfo {year} {2006}{\natexlab{b}})}\BibitemShut
  {NoStop}%
\bibitem [{\citenamefont {Meg\'ias}\ and\ \citenamefont
  {Santos}(2021{\natexlab{a}})}]{paperIII}%
  \BibitemOpen
  \bibfield  {author} {\bibinfo {author} {\bibfnamefont {A.}~\bibnamefont
  {Meg\'ias}}\ and\ \bibinfo {author} {\bibfnamefont {A.}~\bibnamefont
  {Santos}},\ }\href@noop {} {\bibinfo {title} {Translational and angular
  velocity cumulants in granular gases of inelastic and rough hard disks or
  spheres}}\ \bibinfo {howpublished} {(unpublished)} \BibitemShut {NoStop}%
\bibitem [{\citenamefont {Brilliantov}\ \emph {et~al.}(2007)\citenamefont
  {Brilliantov}, \citenamefont {P\"oschel}, \citenamefont {Kranz},\ and\
  \citenamefont {Zippelius}}]{BPKZ07}%
  \BibitemOpen
  \bibfield  {author} {\bibinfo {author} {\bibfnamefont {N.~V.}\ \bibnamefont
  {Brilliantov}}, \bibinfo {author} {\bibfnamefont {T.}~\bibnamefont
  {P\"oschel}}, \bibinfo {author} {\bibfnamefont {W.~T.}\ \bibnamefont
  {Kranz}},\ and\ \bibinfo {author} {\bibfnamefont {A.}~\bibnamefont
  {Zippelius}},\ }\bibfield  {title} {\bibinfo {title} {Translations and
  rotations are correlated in granular gases},\ }\href
  {https://doi.org/10.1103/PhysRevLett.98.128001} {\bibfield  {journal}
  {\bibinfo  {journal} {Phys. Rev. Lett.}\ }\textbf {\bibinfo {volume} {98}},\
  \bibinfo {pages} {{128}{001}} (\bibinfo {year} {2007})}\BibitemShut {NoStop}%
\bibitem [{\citenamefont {Santos}\ \emph {et~al.}(2011)\citenamefont {Santos},
  \citenamefont {Kremer},\ and\ \citenamefont {dos Santos}}]{SKS11}%
  \BibitemOpen
  \bibfield  {author} {\bibinfo {author} {\bibfnamefont {A.}~\bibnamefont
  {Santos}}, \bibinfo {author} {\bibfnamefont {G.~M.}\ \bibnamefont {Kremer}},\
  and\ \bibinfo {author} {\bibfnamefont {M.}~\bibnamefont {dos Santos}},\
  }\bibfield  {title} {\bibinfo {title} {Sonine approximation for collisional
  moments of granular gases of inelastic rough spheres},\ }\href
  {https://doi.org/10.1063/1.3558876} {\bibfield  {journal} {\bibinfo
  {journal} {Phys. Fluids}\ }\textbf {\bibinfo {volume} {23}},\ \bibinfo
  {pages} {{030}{604}} (\bibinfo {year} {2011})}\BibitemShut {NoStop}%
\bibitem [{\citenamefont {{Vega Reyes}}\ \emph
  {et~al.}(2014{\natexlab{a}})\citenamefont {{Vega Reyes}}, \citenamefont
  {Santos},\ and\ \citenamefont {Kremer}}]{VSK14}%
  \BibitemOpen
  \bibfield  {author} {\bibinfo {author} {\bibfnamefont {F.}~\bibnamefont
  {{Vega Reyes}}}, \bibinfo {author} {\bibfnamefont {A.}~\bibnamefont
  {Santos}},\ and\ \bibinfo {author} {\bibfnamefont {G.~M.}\ \bibnamefont
  {Kremer}},\ }\bibfield  {title} {\bibinfo {title} {Role of roughness on the
  hydrodynamic homogeneous base state of inelastic spheres},\ }\href
  {https://doi.org/10.1103/PhysRevE.89.020202} {\bibfield  {journal} {\bibinfo
  {journal} {Phys. Rev. E}\ }\textbf {\bibinfo {volume} {89}},\ \bibinfo
  {pages} {{020}{202}(R)} (\bibinfo {year} {2014}{\natexlab{a}})}\BibitemShut
  {NoStop}%
\bibitem [{\citenamefont {{Vega Reyes}}\ \emph
  {et~al.}(2014{\natexlab{b}})\citenamefont {{Vega Reyes}}, \citenamefont
  {Santos},\ and\ \citenamefont {Kremer}}]{VSK14b}%
  \BibitemOpen
  \bibfield  {author} {\bibinfo {author} {\bibfnamefont {F.}~\bibnamefont
  {{Vega Reyes}}}, \bibinfo {author} {\bibfnamefont {A.}~\bibnamefont
  {Santos}},\ and\ \bibinfo {author} {\bibfnamefont {G.~M.}\ \bibnamefont
  {Kremer}},\ }\bibfield  {title} {\bibinfo {title} {Properties of the
  homogeneous cooling state of a gas of inelastic rough particles},\ }\href
  {https://doi.org/10.1063/1.4902634} {\bibfield  {journal} {\bibinfo
  {journal} {AIP Conf. Proc.}\ }\textbf {\bibinfo {volume} {1628}},\ \bibinfo
  {pages} {494} (\bibinfo {year} {2014}{\natexlab{b}})}\BibitemShut {NoStop}%
\bibitem [{\citenamefont {{Vega Reyes}}\ and\ \citenamefont
  {Santos}(2015)}]{VS15}%
  \BibitemOpen
  \bibfield  {author} {\bibinfo {author} {\bibfnamefont {F.}~\bibnamefont
  {{Vega Reyes}}}\ and\ \bibinfo {author} {\bibfnamefont {A.}~\bibnamefont
  {Santos}},\ }\bibfield  {title} {\bibinfo {title} {Steady state in a gas of
  inelastic rough spheres heated by a uniform stochastic force},\ }\href
  {https://doi.org/10.1063/1.4934727} {\bibfield  {journal} {\bibinfo
  {journal} {Phys. Fluids}\ }\textbf {\bibinfo {volume} {27}},\ \bibinfo
  {pages} {{113}{301}} (\bibinfo {year} {2015})}\BibitemShut {NoStop}%
\bibitem [{\citenamefont {Lutsko}(2001)}]{L01}%
  \BibitemOpen
  \bibfield  {author} {\bibinfo {author} {\bibfnamefont {J.~F.}\ \bibnamefont
  {Lutsko}},\ }\bibfield  {title} {\bibinfo {title} {Model for the atomic-scale
  structure of the homogeneous cooling state of granular fluids},\ }\href
  {https://doi.org/10.1103/PhysRevE.63.061211} {\bibfield  {journal} {\bibinfo
  {journal} {Phys. Rev. E}\ }\textbf {\bibinfo {volume} {63}},\ \bibinfo
  {pages} {061211} (\bibinfo {year} {2001})}\BibitemShut {NoStop}%
\bibitem [{\citenamefont {Luding}\ and\ \citenamefont {McNamara}(1998)}]{LM98}%
  \BibitemOpen
  \bibfield  {author} {\bibinfo {author} {\bibfnamefont {S.}~\bibnamefont
  {Luding}}\ and\ \bibinfo {author} {\bibfnamefont {S.}~\bibnamefont
  {McNamara}},\ }\bibfield  {title} {\bibinfo {title} {How to handle the
  inelastic collapse of a dissipative hard-sphere gas with the {TC} model},\
  }\href {https://doi.org/10.1007/s100350050017} {\bibfield  {journal}
  {\bibinfo  {journal} {Granul. Matter}\ }\textbf {\bibinfo {volume} {1}},\
  \bibinfo {pages} {113} (\bibinfo {year} {1998})}\BibitemShut {NoStop}%
\bibitem [{\citenamefont {Hosoya}\ \emph {et~al.}(2004)\citenamefont {Hosoya},
  \citenamefont {Buchert},\ and\ \citenamefont {Morita}}]{HBM04}%
  \BibitemOpen
  \bibfield  {author} {\bibinfo {author} {\bibfnamefont {A.}~\bibnamefont
  {Hosoya}}, \bibinfo {author} {\bibfnamefont {T.}~\bibnamefont {Buchert}},\
  and\ \bibinfo {author} {\bibfnamefont {M.}~\bibnamefont {Morita}},\
  }\bibfield  {title} {\bibinfo {title} {Information entropy in cosmology},\
  }\href {https://doi.org/10.1103/PhysRevLett.92.141302} {\bibfield  {journal}
  {\bibinfo  {journal} {Phys. Rev. Lett.}\ }\textbf {\bibinfo {volume} {92}},\
  \bibinfo {pages} {141302} (\bibinfo {year} {2004})}\BibitemShut {NoStop}%
\bibitem [{\citenamefont {Akerblom}\ and\ \citenamefont
  {Cornelissen}(2012)}]{AC12}%
  \BibitemOpen
  \bibfield  {author} {\bibinfo {author} {\bibfnamefont {N.}~\bibnamefont
  {Akerblom}}\ and\ \bibinfo {author} {\bibfnamefont {G.}~\bibnamefont
  {Cornelissen}},\ }\bibfield  {title} {\bibinfo {title} {Relative entropy as a
  measure of inhomogeneity in general relativity},\ }\href
  {https://doi.org/10.1063/1.3675440} {\bibfield  {journal} {\bibinfo
  {journal} {J. Math. Phys.}\ }\textbf {\bibinfo {volume} {53}},\ \bibinfo
  {pages} {012502} (\bibinfo {year} {2012})}\BibitemShut {NoStop}%
\bibitem [{\citenamefont {{Shannon}}(1948)}]{S48}%
  \BibitemOpen
  \bibfield  {author} {\bibinfo {author} {\bibfnamefont {C.~E.}\ \bibnamefont
  {{Shannon}}},\ }\bibfield  {title} {\bibinfo {title} {A mathematical theory
  of communication},\ }\href
  {https://doi.org/10.1002/j.1538-7305.1948.tb01338.x} {\bibfield  {journal}
  {\bibinfo  {journal} {Bell Syst. Tech. J.}\ }\textbf {\bibinfo {volume}
  {27}},\ \bibinfo {pages} {379} (\bibinfo {year} {1948})}\BibitemShut
  {NoStop}%
\bibitem [{\citenamefont {Schneider}\ \emph {et~al.}(2021)\citenamefont
  {Schneider}, \citenamefont {Musiolik}, \citenamefont {Kollmer}, \citenamefont
  {Steinpilz}, \citenamefont {Kruss}, \citenamefont {Jungmann}, \citenamefont
  {Demirci}, \citenamefont {Teiser},\ and\ \citenamefont {Wurm}}]{SMKSKJDTW21}%
  \BibitemOpen
  \bibfield  {author} {\bibinfo {author} {\bibfnamefont {N.}~\bibnamefont
  {Schneider}}, \bibinfo {author} {\bibfnamefont {G.}~\bibnamefont {Musiolik}},
  \bibinfo {author} {\bibfnamefont {J.~E.}\ \bibnamefont {Kollmer}}, \bibinfo
  {author} {\bibfnamefont {T.}~\bibnamefont {Steinpilz}}, \bibinfo {author}
  {\bibfnamefont {M.}~\bibnamefont {Kruss}}, \bibinfo {author} {\bibfnamefont
  {F.}~\bibnamefont {Jungmann}}, \bibinfo {author} {\bibfnamefont
  {T.}~\bibnamefont {Demirci}}, \bibinfo {author} {\bibfnamefont
  {J.}~\bibnamefont {Teiser}},\ and\ \bibinfo {author} {\bibfnamefont
  {G.}~\bibnamefont {Wurm}},\ }\bibfield  {title} {\bibinfo {title}
  {Experimental study of clusters in dense granular gas and implications for
  the particle stopping time in protoplanetary disks},\ }\href
  {https://doi.org/10.1016/j.icarus.2021.114307} {\bibfield  {journal}
  {\bibinfo  {journal} {Icarus}\ }\textbf {\bibinfo {volume} {360}},\ \bibinfo
  {pages} {114307} (\bibinfo {year} {2021})}\BibitemShut {NoStop}%
\bibitem {note_21_07_1}%
  \BibitemOpen
  \href@noop {} {\bibinfo {title} {See  {S}upplemental {M}aterial at
  \href{http://link.aps.org/supplemental/10.1103/PhysRevE.104.034902}{http://link.aps.org/supplemental/10.1103/PhysRevE.104.034902}
  for videos with snapshots from $s=0$ to $s=1000$.}}\BibitemShut {Stop}%
\bibitem [{\citenamefont {Mitrano}\ \emph {et~al.}(2011)\citenamefont
  {Mitrano}, \citenamefont {Dahl}, \citenamefont {Cromer}, \citenamefont
  {Pacella},\ and\ \citenamefont {Hrenya}}]{MDCPH11}%
  \BibitemOpen
  \bibfield  {author} {\bibinfo {author} {\bibfnamefont {P.~P.}\ \bibnamefont
  {Mitrano}}, \bibinfo {author} {\bibfnamefont {S.~R.}\ \bibnamefont {Dahl}},
  \bibinfo {author} {\bibfnamefont {D.~J.}\ \bibnamefont {Cromer}}, \bibinfo
  {author} {\bibfnamefont {M.~S.}\ \bibnamefont {Pacella}},\ and\ \bibinfo
  {author} {\bibfnamefont {C.~M.}\ \bibnamefont {Hrenya}},\ }\bibfield  {title}
  {\bibinfo {title} {Instabilities in the homogeneous cooling of a granular
  gas: A quantitative assessment of kinetic-theory predictions},\ }\href
  {https://doi.org/10.1063/1.3633012} {\bibfield  {journal} {\bibinfo
  {journal} {Phys. Fluids}\ }\textbf {\bibinfo {volume} {23}},\ \bibinfo
  {pages} {093303} (\bibinfo {year} {2011})}\BibitemShut {NoStop}%
\end{thebibliography}

%\end{document}


\end{document}